\documentclass[lettersize,journal]{IEEEtran}  

\usepackage{tikz}
\usetikzlibrary{shapes.geometric}
\usetikzlibrary{decorations.pathreplacing}
\usetikzlibrary{positioning, calc}
\usetikzlibrary{patterns}
\usetikzlibrary{arrows.meta} 

\newcommand{\review}[1]{{\color{black}#1}}

\usepackage{pgfplots}
\pgfplotsset{compat=1.18}

\usepackage{mathtools}
\usepackage{threeparttable}
\usepackage{amssymb}
\usepackage{bbm}
\usepackage{hhline}
\usepackage{xcolor}

\usepackage{stfloats}
\usepackage{url}

\usepackage{balance}

\ifCLASSINFOpdf
\else
\fi
\usepackage{amsmath}
\pgfplotsset{compat=1.14} 

\begin{document}

\title{From Network Experience to Subscriber Retention: An Explainable AI Framework for Mobile Operators}
\author{Faris~B.~Mismar,~\IEEEmembership{Senior~Member,~IEEE}, 
    Abdol Saleh, 
    Ivan Maxmillian Putra Pasaribu, %
    and Suhelmy Syaifuddin %
\thanks{Faris B. Mismar (\textit{corresponding author}) and Abdol Saleh are with Nokia Bell Labs Consulting; Ivan Maxmillian Putra Pasaribu and Suhelmy Syaifuddin are independent consultants.}
}%

\maketitle

\begin{abstract}
This paper presents an explainable AI-driven framework for predicting net churn among prepaid mobile subscribers. The framework formulates churn prediction as a machine learning task mapped onto an industry reference architecture.  Moreover, it introduces a quantile-based feature densification transform that reconciles learning features of unequal length and granularity into a homogeneous design matrix, \review{evaluates a label-aware truncation diagnostic that sets aside low-confidence subscribers,} and generates automated English-language insights describing the impact of each feature on churn propensity.  We implement and validate the framework on real production data from a globally leading telco with tens of millions of prepaid subscribers spanning approximately $800$ learning features and over $250$~TB of commercial, experience, and network data. \review{As a diagnostic, setting aside roughly $3$\% of the subscribers, selected using the true labels, lifts $F_1$ on the retained subscribers to the target.  This is not an achievable operating performance on all subscribers.  In our data and feature set, aggregated radio access network counters carried less predictive signal than commercial and experience indicators, despite their greater representation.  The quality of experience index, which is derived from network measurements, ranked second in importance.  Our results also suggest} that churn risk rises as subscriber value declines, with high-value segments disproportionately exposed to gradual value leakage rather than outright attrition.
\end{abstract}


\maketitle

\section{Introduction}
\label{sec:introduction}
The surge in smartphone subscriptions has been fueled by the widespread availability of technology, reduced device costs, application-centric digital ecosystems, and the growing reliance on mobile connectivity for all facets of life. Telecommunication operators (or ``telcos'') are increasingly encountering flat or saturated mobile penetration levels (i.e., the percentage of a population that has active subscriptions).  In this environment, sustaining customer loyalty becomes a strategic priority and a core competitive capability for telcos.  Telcos continue to simultaneously operate several generations of wireless communications using the second (2G), third (3G), and fourth (4G) generation of wireless communications.  Deployments of the fifth generation of wireless communications (5G) continue due to demands for much higher data rates and lower latency than what 4G offered.  A drive to modernize networks and repurpose spectrum motivates telcos to phase out older technologies such as 2G and 3G \cite{digi}.  Amidst this all, subscribers may choose to leave the telco, or \textit{churn}.

While churn in post-paid subscription is defined by events such as contract cancellation, we focus on the churn definition for pre-paid subscription, which is not as clear.  \review{In this context, churn is defined as a subscriber becoming non–revenue generating.  A prepaid subscriber is a gross churner if no account top-ups occur for $180$ consecutive days; a net churner is a gross churner who does not resume payment activity before the $180$ days expire.  The \textit{threshold period} (TP) of three months is the prediction horizon, that is, the lead time available to the operator before the $180$ days are exhausted as in Fig.~\ref{fig:timeline}.}  Churn can happen across all wireless communication technologies alike.

Artificial intelligence (AI) and machine learning (ML) support data-driven decision processes by extracting patterns from large volumes of operational and customer data (i.e., big data). By learning from subscriber behavior, usage trends, and service interactions, churn prediction models can strengthen customer relationship management, inform smarter pricing and billing strategies, and optimize the allocation of constrained operational resources in advance---ultimately improving retention and safeguarding revenue streams.

Explainable AI introduces a set of techniques that enable human users to understand, interpret, and trust the decisions produced by the ML model.  It addresses the limitation of not knowing why a particular prediction was made by providing interpretable explanations, such as feature importance.

\subsection{Related Work}
There is no shortage of prior work on churn prediction whether it is a classical statistical model or a modern explainable AI pipeline.  However, the large majority of recent studies validate the performance of their models on small, static, and public datasets.  The canonical example is the ``Telco customer churn: IBM dataset'' Kaggle dataset.  Using this dataset, several recent Shapley additive explanations (SHAP)-based frameworks combining ensemble or boosting classifiers with post-hoc explanation have been used \cite{asif2025,elattar2026,rasool2025}.  However, they operate at a scale several orders of magnitude smaller than an operational deployment and do not produce automated and decision-ready explanation artifacts that this paper offers. 

An attempt to use deep learning to capture temporal usage patterns was covered in \cite{churnnet2024,joy2024}.  In spite of the use of time-series datasets, there is a requirement for these datasets to have similar length.  However, in practice not all telco data is sampled at the same intervals.  Our approach deals with this practical problem and deals with multiple domains of data at a telco scale.

\review{Relative to the studies in \cite{asif2025,elattar2026,rasool2025,churnnet2024,joy2024}, our setting differs in: (1) scale (i.e., tens of millions of subscribers) in contrast to the public benchmark datasets of \cite{asif2025,elattar2026,rasool2025}), (2) the heterogeneity of the inputs (i.e., commercial, experience, and network data of unequal length and granularity), and (3) in the treatment of low-confidence cases.}

Within telco operations specifically, prior work on unsupervised performance self-diagnosis has shown that live RAN counters can be mined for anomaly signals in real time \cite{9500204}, but this line of work has not been connected to subscriber-level churn outcomes or explainability.


In this paper, we propose an explainable AI-driven framework designed to forecast net churn in advance in realistic telcos, enabling proactive and targeted intervention strategies and deriving insights about the churn at large.

The main contributions of this paper can be summarized as:
\begin{enumerate}
\review{\item Demonstrating how the idea of density transforms of learning features of different lengths can be applied toward creating a design matrix; this is the primary methodological contribution.}
\review{\item Evaluating a label-aware truncation diagnostic that removes the subscribers the model ranks least confidently in their true class.}
\review{\item Formulating the churn prediction problem as a machine learning task and showing how it aligns with industry reference architectures.}
\review{\item Introducing customer lifetime value as a measure of subscriber value to the telco, and showing that churn risk rises as subscriber value declines, with high-value segments disproportionately exposed to gradual value leakage rather than outright churn.}
\review{\item Proposing an approach for influential feature explainability including generation of English-based insights describing impact on churn propensity.} 
\end{enumerate}%

\begin{figure}[!t]
\centering
\resizebox{0.48\textwidth}{!}{\review{%
\begin{tikzpicture}[>=Latex, font=\small]
\draw[->, line width=1pt] (-0.3,0) -- (9.8,0) node[below left] {time};
\draw[fill=blue!15] (0,0.15) rectangle (6,0.75);
\node at (3,0.45) {observation window ($6$ months)};
\node[font=\footnotesize] at (3,1.05) {features computed here only};
\draw[fill=orange!25] (6,0.15) rectangle (9,0.75);
\node at (7.5,0.45) {horizon ($3$ months)};
\node[font=\footnotesize] at (7.5,1.05) {operator lead time};
\draw[line width=1.2pt] (6,-0.25) -- (6,1.3);
\node[above] at (6,1.3) {prediction cutoff};
\draw[<->, line width=0.9pt] (3,-0.7) -- (9,-0.7);
\node[below] at (6,-0.75) {$180$ consecutive days without top-ups};
\draw[dashed] (3,-0.7) -- (3,0);
\draw[dashed] (9,-0.7) -- (9,0.15);
\end{tikzpicture}%
}}%
\caption{\review{Timeline of the observation window, prediction cutoff, prediction horizon, and the $180$-day period.}}
\label{fig:timeline}
\end{figure}

\section{Challenges}\label{sec:challenges} 

Telcos striving to construct a churn model using an explainable AI framework  may face several challenges.  These challenges can be attributed to data, modeling (with machine learning operations (MLOps) pipelines being an example), and business adoption.  In this section we talk about these challenges in adequate detail.

\vspace*{-0.5em}
\subsection{Data challenges}
\subsubsection{Data silos} Data in telcos spans several organizations and domains.  Each one of these may have their own data, which often sits in isolated systems creating a ``siloed'' dataset.  As a result it becomes challenging to build a unified subscriber life cycle view.

\subsubsection{Churn definition} The definition of a churning subscriber depends on several aspects such as pre-paid vs post-paid, period since last activity, and value leakage (i.e., moving to a lower tariff) vs churning out (i.e., leaving the telco).

\subsubsection{`Perfect' explainability} Given the many reasons why a subscriber can churn, it can be nearly impossible to capture all the data features that perfectly explain churn; thus the churn label is likely to be \textit{noisy}.  This is true because churn is associated with a set of limited service touchpoints providing limited observability.  This noisy label can make the prediction performance of the model poor---due to overlapping classes---rendering the model unusable.

\subsubsection{Data aggregation} Some learning features are collected per subscribers; others can be collected per serving base station.   These aggregation granularities are rigid and cannot be modified.  Moreover, data features can have heterogeneous temporal properties (e.g., collected at different time intervals).  As a result creating a key that can be used to join the features into forming a ``design matrix'' \cite{goodfellow} can be challenging.

\subsubsection{Regulatory challenges} Some telcos are bound by regulations.  For example, data is prohibited from leaving the sovereign land where the telco is based. Challenges intensify when the data engineering and data
science teams are located in different geographies. Given the scale and sensitivity of the data described in this paper, ensuring that subscriber and site identifiers are properly hashed, that no directly identifying attributes (e.g., name, address, phone number) are used as learning features, and that access to the resulting design matrix is governed by role-based controls becomes a nontrivial engineering burden layered on top of the modeling task itself.

\vspace*{-1em}
\subsection{Modeling challenges}
\subsubsection{Balanced learning}
Highly imbalanced datasets are common in churn prediction because the number of churners is typically much smaller than the number of retained subscribers.  While many ML algorithms can accommodate class imbalance through  biasing the positive class weight or resampling, extreme imbalance may still degrade model stability and calibration.

\subsubsection{Null ratios and data drift} Another important practical problem comes from null ratios and data drift.   \review{The null ratio computed per learning feature $\mathbf{x}\in\mathbb{R}^{M} \coloneqq \{x_i\}_{i=1}^{M}$, a column vector of the design matrix, is
\begin{equation}
    r_\text{null}(\mathbf{x})\coloneqq \frac{1}{M} \sum_{i=1}^{M} \mathbbm{1}[x_i\;\text{is null}],
\end{equation}
where $M$ is the number of subscribers in the dataset being evaluated and $x_i$ is the value for the $i$-th subscriber, taken from the training design matrix to compute the baseline null ratio and from the inference design matrix to compute the inference null ratio.} Data drift is the change in the underlying probability distribution of a given learning feature over a certain time window (i.e., non-stationarity).  It can be measured using statistical quantities such as the population stability index (PSI).  PSI captures shifts in feature distributions by comparing the probability distributions of training and inference data; it quantifies drift through the weighted difference of these distributions and their logarithmic ratios.  Formally:
\begin{equation}
    \text{PSI} \coloneqq\sum_{j=1}^k (p_j - q_j)\ln \left ( \frac{p_j}{q_j} \right ),
\end{equation}
\review{where $k=10$ is the number of decile bins defined on the training data, and $p_j$ ($q_j$) is the proportion of the inference (training) data in the $j$-th bin.}

In several cases, lack of representative data from all touchpoints results in large missing values and forces some features to be dropped.   Drift can happen due to many reasons such as: price changes, plan changes, new technology rollout, new site build, and campaigns from the competition.  Evolving subscriber dynamics can also be a factor: While some of them are well understood, others remain latent and surface indirectly as noise in the data. Known \review{predictive contributors} include the onboarding of new subscribers with preferences that differ from historical norms, changes in behavior among existing subscribers, environmental and contextual factors, and social influences such as peer, family, or community effects.  Additionally, drift can happen due to \textit{windowing}.  Windowing uses the TP to construct the training data.   For instance, if the TP spans three months (e.g., January through March of the current year), the training features must be derived from an earlier observation window, such as July through December of the previous year.  Because of the windowing, training data is constructed using a fixed observation window per subscriber, while the churn label is defined as a forward-looking indicator within the TP.  During operation, inference is performed on a recurring monthly basis using the same trained model, with each cycle relying on newly transformed monthly data distributions. Throughout this process, both the model prediction scores and indicators of data drift are continuously monitored.  This introduces an age-of-information effect \cite{5984917}: as the input data becomes increasingly distant from the data used during training, model performance may degrade.

\subsubsection{Model explainability} Telco decision makers look for actionable insights to improve the business.  Explainability of a model to address this need is to translate feature importances into specific feature-driven insights that can be used for churn mitigation actions.  In this case, the framework provides explainability in terms of human-language insights and customer life-time value.

\subsubsection{Productionization and time horizon} Transforming a validated model into a deployable, enterprise-grade solution while establishing reliable implementation under comprehensive model life-cycle governance poses a significant challenge for telcos.  This process also encompasses continuous training, systematic validation, and periodic retraining to maintain performance and relevance over time.  We call this validity time the \textit{time horizon} of a model that represents its longevity.

\subsection{Business challenges} 
\subsubsection{Actionability gap} 
This means that a strategy is needed to act upon the insights generated by the churn prediction model.  Often, the concept of uplift (i.e., how many subscribers have been prevented from churning due to an action) is mentioned in this context.  An actionability gap therefore is when an insight cannot be acted upon or does not lead to uplift.
\subsubsection{Organizational silos} 
Telcos are structured around distinct domains (e.g., network engineering, marketing, customer care, and finance).  Each domain has its own performance measures, systems, and leadership.  Therefore, the ownership of a churn prediction model and how teams from the various domains interact to keep the model updated and relevant can be perceived as an extracurricular activity, causing its ownership to become fragmented, ambiguous, and ultimately deprioritized in favor of other domain-specific objectives.

Now that we covered the practical challenges that the construction of a churn model in a telco can face, we next talk about autonomous networks and how the proposed explainable AI-driven framework fits into it.

\section{Autonomous Networks}\label{sec:autonomous}

\begin{figure*}[!t]
\centering
\resizebox{0.85\textwidth}{!}{\begin{tikzpicture}[
  font=\sffamily,
  >=Latex,
  box/.style={draw=black, thick, fill=white, align=center, inner sep=4pt},
  hdrbox/.style={draw=black, thick, fill=gray!8, align=center, inner sep=4pt},
  intel/.style={draw=black, thick, fill=gray!45, align=center, text=black, inner sep=3pt},
  intelfoot/.style={draw=black, thick, fill=gray!75, align=center, text=white, font=\sffamily\bfseries\footnotesize, inner sep=2pt},
  subbox/.style={draw=black, thin, fill=gray!20, align=center, font=\sffamily\scriptsize, inner sep=2.5pt},
  tinybox/.style={draw=black, thin, fill=gray!22, align=center, font=\sffamily\tiny, inner sep=2pt},
  dottedoval/.style={draw=black, dotted, thick, rounded corners=16pt},
  domaindashed/.style={draw=black, dashed, thick, rounded corners=4pt},
  arrowlbl/.style={font=\sffamily\bfseries},
  loopy/.style={draw=black, thick, ->, >=Stealth}
]

\node[hdrbox, minimum width=4.4cm, minimum height=1.0cm, anchor=south] (anconsumers) at (7.0,10.65) {\textbf{AN Consumers}\\ \footnotesize (e.g., external ecosystems)};

\node[box, minimum width=12cm, minimum height=1.4cm, anchor=north west] (bizops) at (1.0,10.2) {};
\node[box, minimum width=12cm, minimum height=1.4cm, anchor=north west] (svcops) at (1.0,8.25) {};

\node[domaindashed, minimum width=3.6cm, minimum height=2.0cm, anchor=north west] (domA) at (1.0,6.35) {};
\node[domaindashed, minimum width=3.6cm, minimum height=2.0cm, anchor=north west] (domB) at (5.1,6.35) {};

\node[box, minimum width=3.5cm, minimum height=5.75cm, anchor=north west] (knowbox) at (15.2,10.2) {};

\draw[loopy] (anconsumers.south) -- (7.0,10.2);
\node[arrowlbl] at (7.4,10.42) {I1};

\node[anchor=north west, font=\sffamily\bfseries] (biztitle) at ($(bizops.north west)+(0.15,-0.12)$) {Business Operations};

\node[intel, minimum width=2.9cm, minimum height=0.75cm, anchor=north east] (bizintel) at ($(bizops.north east)+(-0.2,-0.2)$) {\footnotesize Business Operations\\ \footnotesize Intelligence};
\node[intelfoot, minimum width=2.9cm, anchor=north] (bizintelfoot) at (bizintel.south) {Local Intelligence};

\path let \p1=(bizops.west) in coordinate (f1line) at (0,\y1);
\draw[thick] (f1line) -- (bizops.west);
\node[arrowlbl, fill=white, inner sep=1pt] (f1) at ($(bizops.west)+(-0.55,0)$) {F1};

\path let \p1=(bizintel.east), \p2=(knowbox.west) in coordinate (k1tip) at (\x2,\y1);
\draw[loopy] (bizintel.east) -- (k1tip);
\node[arrowlbl] at ($(bizintel.east)+(0.75,0.22)$) {K1};

\draw[loopy] (7.0,8.8) -- (7.0,8.25);
\node[arrowlbl] at (7.4,8.53) {I2};

\node[anchor=north west, font=\sffamily\bfseries] (svctitle) at ($(svcops.north west)+(0.15,-0.12)$) {Service Operations};

\node[intel, minimum width=2.9cm, minimum height=0.75cm, anchor=north east] (svcintel) at ($(svcops.north east)+(-0.2,-0.2)$) {\footnotesize Service Operations\\ \footnotesize Intelligence};
\node[intelfoot, minimum width=2.9cm, anchor=north] (svcintelfoot) at (svcintel.south) {Local Intelligence};

\path let \p1=(svcops.west) in coordinate (f2line) at (0,\y1);
\draw[thick] (f2line) -- (svcops.west);
\node[arrowlbl, fill=white, inner sep=1pt] (f2) at ($(svcops.west)+(-0.55,0)$) {F2};

\path let \p1=(svcintel.east), \p2=(knowbox.west) in coordinate (k2tip) at (\x2,\y1);
\draw[loopy] (svcintel.east) -- (k2tip);
\node[arrowlbl] at ($(svcintel.east)+(0.75,0.22)$) {K2};

\draw[loopy] (7.0,6.85) -- (7.0,6.35);
\node[arrowlbl, fill=white, inner sep=1pt] at (7.4,6.65) {I3};

\node[font=\sffamily\bfseries\Huge] at (11.45,5.5) {$\cdots$};

\node[box, minimum width=3.2cm, minimum height=0.9cm, anchor=north west] (resopA) at ($(domA.north west)+(0.2,-0.3)$) {};
\node[anchor=west, font=\sffamily\small, xshift=3pt, text width=1.5cm, align=left] at (resopA.west) {Resource\\Operations};
\node[intel, minimum height=0.5cm, anchor=north east, font=\sffamily\tiny, xshift=-4pt, yshift=-4pt, text width=1.05cm, align=center] (domintelA) at (resopA.north east) {Domain Intelligence};

\path let \p1=(domA.west) in coordinate (f3line) at (0,\y1);
\draw[thick] (f3line) -- (domA.west);
\node[arrowlbl, fill=white, inner sep=1pt] (f3a) at ($(domA.west)+(-0.55,0)$) {F3};

\node[font=\sffamily\footnotesize\itshape, text=black!55, align=center] at ($(domA.south)+(0,0.4)$) {Autonomous Domain A};

\node[box, minimum width=3.2cm, minimum height=0.9cm, anchor=north west] (resopB) at ($(domB.north west)+(0.2,-0.3)$) {};
\node[anchor=west, font=\sffamily\small, xshift=3pt, text width=1.5cm, align=left] at (resopB.west) {Resource\\Operations};
\node[intel, minimum height=0.5cm, anchor=north east, font=\sffamily\tiny, xshift=-4pt, yshift=-4pt, text width=1.05cm, align=center] (domintelB) at (resopB.north east) {Domain Intelligence};


\node[font=\sffamily\footnotesize\itshape, text=black!55, align=center] at ($(domB.south)+(0,0.4)$) {Autonomous Domain B};

\path let \p1=(domB.east), \p2=(knowbox.west) in coordinate (k3tip) at (\x2,\y1);
\draw[loopy] (domB.east) -- (k3tip);
\node[arrowlbl] at ($(domB.east)+(0.75,0.22)$) {K3};

\node[anchor=north, font=\sffamily\bfseries, align=center] (knowtitle) at ($(knowbox.north)+(0,-0.27)$) {Knowledge and\\ Intelligence};
\draw[black, thick] ($(knowbox.north west)+(0,-1.15)$) -- ($(knowbox.north east)+(0,-1.15)$) coordinate[midway] (kdiv1);

\node[subbox, minimum width=2.6cm, minimum height=0.6cm, font=\sffamily\small, below=0.2cm of kdiv1] (appcat) {Application\\Catalog};

\path let \p1=(appcat.south), \p2=(knowbox.north) in coordinate (mtkbanchor) at (\x2,\y1);
\node[subbox, text width=1.5cm, minimum height=0.85cm, font=\sffamily\small, anchor=east, 
yshift=-0.4cm] (modtrain) at (mtkbanchor) {Model\\Training};
\node[subbox, text width=1.5cm, minimum height=0.85cm, font=\sffamily\small, anchor=west, 
yshift=-0.4cm] (knowbase) at (mtkbanchor) {Knowledge\\Base};

\draw[black, thick] ($(modtrain.south west)+(-0.1,-0.1)$) -- ($(knowbase.south east)+(0.1,-0.1)$) coordinate[midway] (kdiv2);

\node[subbox, minimum width=2.6cm, minimum height=0.6cm, font=\sffamily\small, below=0.4cm of kdiv2] (dataserv) {Data Services};

\end{tikzpicture}}%
\caption{Autonomous networks reference architecture as defined in \cite{tmforum2022autonomous}.}%
\label{fig:an}
\end{figure*}%

Autonomous networks (ANs) as shown in Fig.~\ref{fig:an} are networks that are intent-driven, context-aware, and guided by knowledge and intelligence, with autonomous domains serving as the modular building blocks operators assemble into hierarchical, self-governing networks \cite{tmforum2022autonomous}. The TM Forum's AN reference architecture is the industry's common language for how telcos structure closed-loop, self-governing network operations.  This architecture comprises three horizontal operations layers between the outside world (including the subscribers) and the physical network.  These layers are:

\begin{itemize}
    \item Business Operations (F1) sits at the top and handles customer-facing functions (e.g., digital storefronts, and portals).
    \item Service Operations (F2) sits underneath and covers planning, deployment, maintenance and optimization.
    \item Resource Operations (F3) sits at the bottom and directly over the network elements and site infrastructure.
\end{itemize}

What separates this from ordinary automation is that every layer carries its own local intelligence, and each Resource Operations instance lives inside an autonomous domain (ADs).  An AD is a self-contained unit that senses, reasons, decides and acts in a loop. ADs comprise intent management functions that turn stated goals into something machine-actionable, a domain intelligence block, and a control loop that cycles through awareness, analysis, decision, and execution.

Running alongside all three operations layers is a separate Knowledge and Intelligence plane, supplying the application catalog, shared knowledge base, model training and data services that each domain's local intelligence draws on. It functions as the shared reasoning layer that keeps every domain trained and aligned rather than operating in isolation.  Moreover, each layer connects to its neighbors (and to the knowledge and intelligence plane) through a set of standardized interfaces I1, I2, and I3 (K1, K2, and K3). 

\subsection{Mapping churn to autonomous networks}
Churn maps naturally onto this layered structure because subscriber attrition is rarely caused by a single failure at a single layer.  Upon traversing the AN layers it should be clear that at the resource operations layer, autonomous domains continuously sense radio- and core-network level indicators, which we discuss later in Section~\ref{sec:framework}. The service operations attempts to re-route traffic at congestion through optimization and such, but it is the Knowledge and Intelligence plane that turns the various signals of this closed-loop operation into a the single churn risk score. Business Operations then acts on
that shared signal at the point of customer contact, triggering a retention offer through the digital storefront or a portal before the subscriber finally decides to leave. 

Now that we covered the practical challenges telcos can face and motivated churn as a use case for ANs, we next talk about the opportunities that telecommunication operators can gain from having their own churn models in an explainable AI framework.

\section{Opportunities for Telecommunication Operators}\label{sec:opportunities}

This section discusses some of the opportunities enabled by constructing churn models using an explainable AI framework.  These opportunities are improved efficiency, observability, and decision intelligence.

\subsection{Efficiency}
AI can improve the classification performance of churn prediction.  It can also accelerate the analysis and outputs to fraction of the time.  These capabilities allow operators to respond faster to an imminent change in a subscriber behavior.

\subsection{Observability}
A churn model using explainable AI improves the observability task making the model outputs easier to monitor, interpret, and act upon.  Here is how:

\begin{enumerate}
    \item \textbf{Churn propensity}: The model outputs the churn labels $\mathbf{y}\in\{0,1\}_{i=1}^N$ of either retain (``0'') or churn (``1'') and the probability of churn (often called ``churn propensity'' or risk as stated in Section~\ref{sec:autonomous}). This value is a real number that lies between $0$ and $1$.
    \item \textbf{Reasons and explanations}: The framework provides explanations for the predicted churn risk. These explanations reveal the main contributing factors explaining churn at large.  They may include \review{predictive contributors} from across different telco domains such as pricing, usage behavior, network experience, or customer care interactions.  The explanations can also be aggregated to produce insights across customer segments.
\end{enumerate}

\subsection{Decision intelligence}
Informed operational decisions are facilitated by  the churn model ability to provide transparent explanations and proper insights about subscriber retention.  Here is what this means:%
\begin{enumerate}
    \item \textbf{Improved decision transparency}: Model outputs and explanations make predictions easier to understand. This increases confidence in the system across teams.
    \item \textbf{Improved customer retention strategies}: The identified churn \review{predictive contributors} by the order of feature importance help telcos design targeted retention actions.  These actions may include new offerings, service improvements, pricing adjustments, or proactive customer support.
\end{enumerate}%


\section{Proposed Framework}\label{sec:framework} 
After outlining the challenges and opportunities faced by telcos in developing predictive churn models using an explainable AI framework, we now delve into the framework we propose.%

\subsection{Problem statement}
A major global telco serving tens of millions of prepaid subscribers across 2G, 4G, and 5G radio access networks (from several vendors) requires a solution that can (1) use up to $30$\% of the subscriber data of six months of history for training, (2) proactively predict net churners three months ahead of time with a recall of at least $0.7$ and F\textsubscript{1} of at least $0.75$ for the remaining (and unseen) subscriber data (of $70\%$), (3) explain the \review{predictive contributors of} subscriber churn in the network (including the telco domain they belong to) through insights, and 4) be usable (i.e., for inference) for a time horizon of at least three months.

This problem statement focuses exclusively on net churners; \review{a net churner is a gross churner who does not resume payment activity before the $180$ days expire.}  We propose a framework that addresses this problem statement.%

\subsection{Framework elements}
\subsubsection{Platform and governance} Due to the anticipated size of data and in case of sovereignty requirements, a platform as a service (PaaS)  provider can be employed to support scalable data processing and compute requirements through an instance residing in the country.

Subscriber phone numbers and site identifiers were irreversibly hashed before ingestion into the pipeline.  Furthermore, this study was conducted under the operator's existing data governance and access-control authorization, consistent with the data-sovereignty requirements outlined in Section~\ref{sec:challenges}.

\subsubsection{Data sources} Given that telco datasets span several domains and are often siloed as we stated earlier, identifying them is an essential element of the framework:

\begin{enumerate}
    \item\label{item_a} Billing and charging data, including subscriber tenure, tariff plans, and average revenue per user  for the current and prior months.
    \item\label{item_b} Marketing and customer care data, covering exit surveys, trouble tickets, and call center interactions.
    \item Network performance and utilization data, capturing KPIs across 2G, 3G, 4G, and 5G RAN as well as the core network. This includes a composite quality of experience (QoE) index and the subscriber's most frequently served cell site. The QoE index aggregates metrics such as video quality, initial buffering delay, and streaming stalling events.
    \item Geolocation and minimization of drive testing (MDT) data, providing session-level information such as serving cell identifier, device vendor, received signal power, throughput, and latency.
\end{enumerate}%

For analytical clarity, data sources \ref{item_a} and \ref{item_b} are grouped as \textit{commercial data}, the QoE index is categorized as \textit{experience data}, and the remaining radio KPIs, core KPIs, and MDT data are referred to as \textit{network data}. This dataset categorization enables collective assessment of their respective contributions to churn and association to the proper telco domain.   Additionally, metadata to translate the names of features into English (e.g., a lookup table) is required.%

\subsubsection{Data preprocessing} Given that data can have different keys (e.g., at the subscriber level vs at the serving base station level) or heterogeneous temporal properties (e.g., daily vs hourly), a processing step is to construct the design matrix which later serves as an input to the churn prediction model.  More about this is covered in Section~\ref{sec:architecture}.

\subsubsection{Machine learning} The feature densification transform allowed our design matrix to have a homogeneous shape. When combined with the churn prediction labels, the churn prediction task can be formulated as an ML classification task on a supertable as defined in Section~\ref{sec:architecture}. An ML algorithm that works quite well such a supertable for the churn prediction task is the Extreme Gradient Boosting (XGBoost) classifier \cite{xgboost}. XGBoost is particularly well suited for tabular datasets, performs robustly in the presence of missing values---once handled as described earlier---and has consistently achieved top rankings in data mining competitions due to its strong predictive performance.  Moreover, XGBoost can create explainable AI models due to its ability to provide feature importance.  To train the model successfully, an effective churn prediction model benefits strongly from a balanced dataset, in which the churn and non-churn classes are represented in equal or near-equal proportions.

\subsubsection{Truncation heuristic} We have established that churn data can be noisy, which is one of the major challenges preventing the performance of a predictive model from meeting the desired classification performance targets.  \review{Since the model provides a probability of class membership for each subscriber, the subscribers on which it is least confident are set aside with a label of \texttt{NaN}, effectively introducing a third category for our classifier (i.e., churned, retained, or uncertain) or $\tilde{\mathbf{y}} \coloneqq \{0, 1, \mathtt{NaN}\}$.  Let $y_m$ be the true label and $\hat{p}_m$ the predicted churn probability of subscriber $m$.  Define the confidence $c_m = \hat{p}_m$ if $y_m = 1$ and $c_m = 1-\hat{p}_m$ if $y_m = 0$, and sort subscribers by descending $c_m$.  With $\hat{y}_m = \mathbbm{1}[\hat{p}_m > 0.5]$, let $F_1(k)$ be the $F_1$ score over the first $k$ subscribers.  The retained set is the first $k^\star = \max\{k : F_1(k) \geq 0.75\}$ subscribers; all others are uncertain and receive $\tilde{y}_m = \texttt{NaN}$.}

\review{The truncation selects subscribers using the true labels, so it cannot be applied at deployment time unless we leverage the stability of the probability in features as measured by the drift defined in Section~\ref{sec:challenges}.  In this case, we flag subscribers whose scores fall in a band around the decision threshold.  Flagged subscribers would be either rescored in the near future or referred to a designated telco department for further action.}%

\subsubsection{Maintenance}
Due to the change in subscriber dynamics (evident in the PSI drift), non-stationarity, null ratios, and windowing effect, the ML model performance is expected to degrade over time.   In this context, changes in two maintenance measures (i.e., the null ratio and PSI variations) are monitored periodically as part of the MLOps efforts. Any \textit{significant} change in the null ratio between the training and any of the future inference datasets or the PSI exceeding a given threshold, combined with the model performance falling below a predefined threshold, indicate the need for model retraining as part of the model maintenance.%

\begin{figure*}[!t]
\centering
\includegraphics[width=0.8\textwidth]{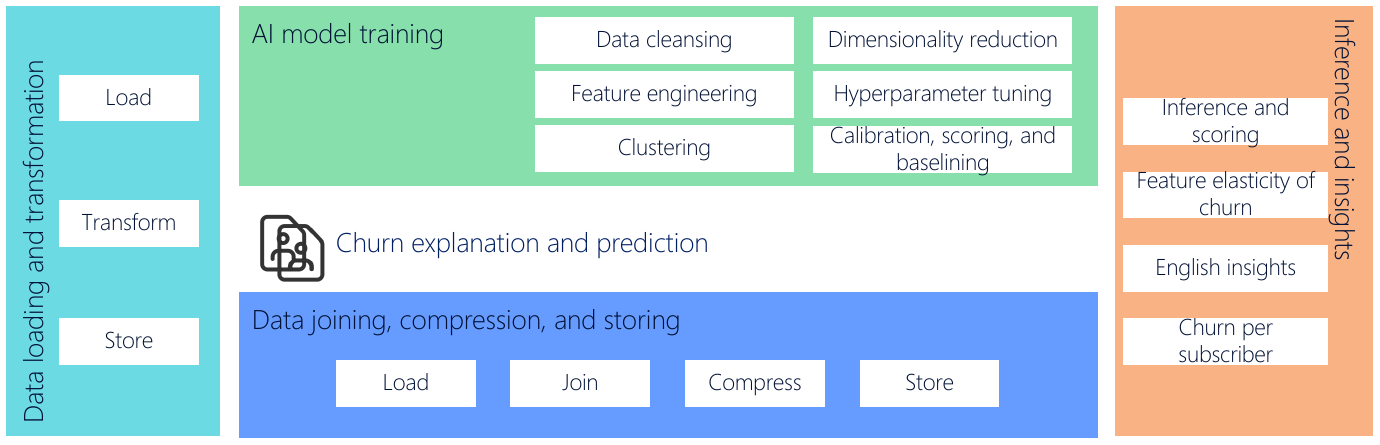}
\caption{Modular architecture for churn explanation and prediction using machine learning.}
\label{fig:mlarchitecture}
\end{figure*}%

\subsection{Value segments and churn}
A subscriber's financial value to telco is defined by their profit margin contribution, commonly referred to as customer life-time value (CLV). This measure is derived from the monthly revenue, usage and other transactional costs, and the total number of months the user remains active. Throughout a subscriber's life cycle, the distribution of this CLV by subscribers and across the user base may vary, appearing skewed to the left, right, or remaining centered depending on the overall health of the portfolio. To capture the shifting value of subscriber equity, telcos can analyze subscriber cohorts using sliding windows. As illustrated in Fig.~\ref{fig:micro_seg}, this approach tracks dynamics ranging from the entry of new subscribers to their specific movements across cohorts with transition probabilities resembling a Markov chain implementing a state transition matrix of various states (or subscriber microsegments). Formally:

\begin{equation}
 \mathbf{C} = \begin{bmatrix}
    p_{1,1} & p_{1,2} & \cdots & p_{1,R} \\
    p_{2,1} & p_{2,2} & \cdots & p_{2,R} \\
    \vdots & \vdots & \ddots & \vdots \\
    p_{R,1} & p_{R,2} & \cdots & p_{R,R}
\end{bmatrix}\in\mathbb{R}^{R\times R},
\end{equation}
where $1$ is the microsegment of the highest value and $R$ is that of the lowest value. What remains is the computation of the various probabilities from the primary predictors of churn. Each predictor is divided
into a predefined number of levels, and the multiplication of these levels determines the total number of distinct value segments. Mathematically:

\begin{equation}
    R = \prod_{e=1}^{E} L_e,
    \label{eq:levels}
\end{equation}
where $E$ is the number of predictors and $L_e$ is the number of levels assigned to predictor $e$. The total number of segments can be obtained from the number of predictors and levels per segments through combinatorics.

The transition probability $p_{i,j}$ is then computed as the empirical frequency with which subscribers in segment $i$ at time $t$ are observed in segment $j$ at time $t+1$:

\begin{equation}
    p_{i,j} = \frac{N_{i,j}}{N_i},
\end{equation}
where $N_{i,j}$ is the number of subscribers transitioning from segment $i$ to segment $j$, and $N_i = \sum_{j=1}^{R} N_{i,j}$.

The total margin contribution to the telco fluctuates based on these internal equity shifts; overall value rises as higher-value microsegments move to the left and diminishes as the density of subscribers shifts right toward lower-value segments. Ultimately, the service life cycle concludes with service termination, or churn, which can occur within any cohorts regardless of the subscriber's value tier. To navigate this complex environment, telcos can leverage AI-based predictive models discussed in this paper.  Based on the churn propensity values, these models can predict tendencies to upgrades, downgrades, and terminations, allowing the telco to calculate the expected impact on its equity (i.e., gross profit margin) and strategic objectives.  The value segments are constructed using the primary churn predictors, as we show in Section~\ref{sec:results}.%
\begin{figure}[!t]
\centering
\resizebox{0.45\textwidth}{!}{\definecolor{PANTONE3015C}{HTML}{00629B}
\begin{tikzpicture}[
  >=Stealth,
  every node/.style={font=\sffamily},
  segnode/.style={circle, draw=black, minimum size=1.4cm, align=center, font=\sffamily\bfseries},
  bignode/.style={ellipse, draw=black, minimum width=2.6cm, minimum height=1.4cm, align=center, font=\sffamily\bfseries},
  newarrow/.style={-{Stealth[length=2mm]}, PANTONE3015C, line width=0.6pt},
  upgradearrow/.style={-{Stealth[length=2mm]}, black, line width=1pt},
  downgradearrow/.style={-{Stealth[length=2mm]}, PANTONE3015C, line width=1pt, dashed},
  churnarrow/.style={-{Stealth[length=2.6mm]}, black, line width=1.8pt},
  legendline/.style={line width=1.4pt}
]
\node[bignode] (newsub) at (5,9) {New Sub};
\node[bignode, fill=gray!45] (churn) at (5,2) {Churned Sub};
\node[segnode] (Sone)    at (0,5.5)  {$1$};
\node[segnode] (Sdotsa)  at (2.5,5.5) {$\cdots$};
\node[segnode] (Sj)      at (5,5.5)  {$j$};
\node[segnode] (Sk)      at (7.5,5.5) {$k$};
\node[segnode] (SR)      at (10,5.5) {$R$};

\draw[newarrow] (newsub) -- (Sone);
\draw[newarrow] (newsub) -- (Sdotsa);
\draw[newarrow] (newsub) -- (Sj);
\draw[newarrow] (newsub) -- (Sk);
\draw[newarrow] (newsub) -- (SR);
\node[PANTONE3015C, above=1pt] at ($(newsub)!0.45!(Sone)$) {$p_{n,1}$};
\node[PANTONE3015C, above=1pt] at ($(newsub)!0.45!(Sj)$)   {$p_{n,j}$};
\node[PANTONE3015C, above=1pt] at ($(newsub)!0.45!(Sk)$)   {$p_{n,k}$};
\node[PANTONE3015C, above=1pt] at ($(newsub)!0.45!(SR)$)   {$p_{n,R}$};
\draw[upgradearrow] (Sone)   to[bend left=40] (Sdotsa);
\draw[upgradearrow] (Sdotsa) to[bend left=40] (Sj);
\draw[upgradearrow] (Sj)     to[bend left=40] (Sk);
\draw[upgradearrow] (Sk)     to[bend left=40] (SR);
\draw[downgradearrow] (Sdotsa) to[bend left=40] (Sone);
\draw[downgradearrow] (Sj)     to[bend left=40] (Sdotsa);
\draw[downgradearrow] (Sk)     to[bend left=40] (Sj);
\draw[downgradearrow] (SR)     to[bend left=40] (Sk);
\node[black]        at ($(Sj)!0.5!(Sk)+(0,0.95)$)  {$p_{j,k}$};
\node[PANTONE3015C] at ($(Sj)!0.5!(Sk)+(0,-1.05)$) {$p_{k,j}$};
\node[black]        at ($(Sk)!0.5!(SR)+(0,0.95)$)  {$p_{k,R}$};
\node[PANTONE3015C] at ($(Sk)!0.5!(SR)+(0,-1.05)$) {$p_{R,k}$};
\draw[churnarrow] (Sone)   to[bend right=15] node[pos=0.5, left=2pt]  {$p_{1,c}$} (churn);
\draw[churnarrow] (Sdotsa) to[bend right=10] (churn);
\draw[churnarrow] (Sj)     -- node[pos=0.5, right=2pt] {$p_{j,c}$} (churn);
\draw[churnarrow] (Sk)     to[bend left=10] node[pos=0.5, right=2pt] {$p_{k,c}$} (churn);
\draw[churnarrow] (SR)     to[bend left=15] node[pos=0.5, right=2pt]  {$p_{R,c}$} (churn);
\draw[newarrow, legendline] (1.3,0.9) -- (2.1,0.9);
\node[right, PANTONE3015C] at (2.2,0.9) {New acquisition, $p_{n,i}$};
\draw[upgradearrow, legendline] (5.6,0.9) -- (6.4,0.9);
\node[right, black] at (6.5,0.9) {Upgrade, $p_{i,i+1}$};
\draw[downgradearrow, legendline] (1.3,0.2) -- (2.1,0.2);
\node[right, PANTONE3015C] at (2.2,0.2) {Downgrade, $p_{i+1,i}$};
\draw[churnarrow, legendline] (5.6,0.2) -- (6.4,0.2);
\node[right, black] at (6.5,0.2) {Churn, $p_{i,c}$};
\end{tikzpicture}}%
\caption{Microsegments and dynamics of loyalty and value in the telco subscriber base.}%
\label{fig:micro_seg}
\end{figure}%

\subsection{Value leakage}
Value leakage is the gradual erosion of customer profitability as opposed to the hard and binary nature customer departure.  Value leakage happens while the subscriber remains active.  In telcos, this may manifest as a decline of revenue, reduced usage intensity, plan downgrades, increased discount dependency, or migration to lower-value bundles. Fig.~\ref{fig:micro_seg} provides a schematic representation of value-transition dynamics within the subscriber base. By observing these dynamics over multiple periods, the transition probabilities between value segments are expected to stabilize as the customer base approaches a steady state. These transition probabilities describe how subscribers move between different value segments and, consequently, how the size and composition of each segment evolve over time. A systematic shift of subscribers toward higher-value segments indicates an improvement in the overall profitability of the customer base, whereas movement toward lower-value segments signals value erosion and declining profit potential for the service provider.  \review{The analysis of value leakage in this paper is descriptive and not causal; causal retention recommendations require the uplift or treatment-effect analysis.}

Subscriber value optimization aims to maximize the economic value generated by each subscriber by identifying differences in subscriber value, detecting sources of value leakage, and determining the interventions most likely to improve future value. Value segmentation provides the foundation for this process by grouping subscribers according to their current and potential value, while leakage analysis identifies where avoidable losses in that value occur. The optimization problem therefore extends beyond identifying high- or low-value subscribers: it seeks to determine which subscribers should be targeted, with which intervention, and with what expected incremental value.%
\section{Architecture}\label{sec:architecture}
Our proposed framework entails a reusable modular architecture that streamlines the various elements mentioned in Section~\ref{sec:framework}.  This architecture includes four different blocks as shown in Fig.~\ref{fig:mlarchitecture}. %

\subsection{Data loading and transformation}
This block ingests raw data from the data sources mentioned earlier, typically through database queries written in structured query language (SQL). However, direct SQL-based retrieval can be inefficient when operating on very large datasets. To address this, the query outputs are stored in more efficient formats, such as columnar files (e.g., Parquet).  Executing this block produces multiple files that are stored in a PaaS-based object storage bucket.  \review{The pipeline is batch-oriented: monthly extracts are stored as columnar files and joined offline.}

The commercial and experience datasets have the subscriber identifier as the primary key while the network data has serving site identifier as the primary key.  Joining these two types to form the design matrix requires a new definition: the \textit{favorite} site identifier.  Every subscriber from the subscriber set $\mathcal{M}$ has a favorite site identifier corresponding to the site that where the subscriber attempts the most sessions.  Formally, let $\mathcal{S}$ be the set of sites in the network, then for a subscriber $m\in\mathcal{M}$, the favorite site identifier is: %
\begin{equation}
    s_m^\star = \underset{s\in\mathcal{S}}{\arg\,\max} \sum_\nu A_s(m, \nu),
\end{equation}%
where $A_s(m, \nu)$ is the number of session attempts of the $m$-th subscriber on the $s$-th site over the measurement interval $\nu$.

To address the problem of heterogeneous temporal properties, we propose a transform.  \review{To start, for a given measurement interval and subscriber identifier $m$ define a transformation.  Here $M$ is the number of subscribers, $L$ is the number of learning features, and $d_{m,k}$ is the sample-dependent length of the $k$-th column vector of subscriber $m$.
\begin{equation}
\begin{aligned}
\varphi(\mathbf{x}_{m,k};Q)&\colon \mathbb{R}^{d_{m,k}}\to\mathbb{R}^{1\times Q},\\
\mathbf{x}^{(Q)}_{m,k} &\coloneqq \begin{bmatrix}F^{-1}_{\mathbf{x}_{m,k}}(q_1) & \cdots & F^{-1}_{\mathbf{x}_{m,k}}(q_Q)\end{bmatrix},\\
\mathbf{X}_m &\coloneqq \big[\mathbf{x}^{(Q)}_{m,k}\big]_{k=1}^{L}\in\mathbb{R}^{L\times Q}
\end{aligned}
\label{eq:quantile-map}
\end{equation}
\begin{equation}
\begin{aligned}
\mathbf{X}^{(Q)} &\coloneqq \big[\mathrm{vec}(\mathbf{X}_1);\ \cdots;\ \mathrm{vec}(\mathbf{X}_M)\big]\in\mathbb{R}^{M\times LQ},\\
\mathrm{vec}(\mathbf{X}_m) &\coloneqq \begin{bmatrix}\mathbf{x}^{(Q)}_{m,1} & \cdots & \mathbf{x}^{(Q)}_{m,L}\end{bmatrix}\in\mathbb{R}^{1\times LQ},
\end{aligned}
\label{eq:transformed_matrix}
\end{equation}
where $\mathrm{vec}(\cdot)$ places the rows of $\mathbf{X}_m$ side by side}
and \review{$F_{\mathbf{x}_{m,k}}^{-1}(q)$} is the empirical quantile function of that vector (i.e., inverts the empirical cumulative distribution function).  In other words:
\begin{equation}
F_{\mathbf{x}_{m,k}}^{-1}(q) = \inf\left\{\, t \in \mathbb{R} \Bigm | \frac{1}{d_{m,k}}\sum_{i=1}^{d_{m,k}} \mathbbm{1}\!\left[x_{m,k}[i] \le t\right] \;\ge\; \frac{q}{100} \,\right\},
\end{equation}%
where $d_{m,k}$ is an arbitrary sample-dependent length and $x_{m,k}[i]$ is the $i$-th element of the column vector $\mathbf{x}_{m,k}$ (i.e., before the transformation is applied).

\review{Subscribers with a maximum tenure below $30$ days are tagged as \emph{cold subscribers} and are dropped prior to applying the transform} due to absence of statistical significance of these aggregation statistics.  This transform outlined in \eqref{eq:quantile-map} effectively causes the number of columns to grow $Q$-fold but captures enough variance for the prediction to perform.  To ensure that feature importance remains meaningful, the generated features continue to use the original feature names with a suffix denoting the aggregation method chosen.  We call this step \textit{feature densification} and the new transformed design matrix is denoted $\mathbf{X}^{(Q)}$ as shown in \eqref{eq:transformed_matrix}.

The feature densification data transformation step is then applied across the relevant learning features to ensure the learning features have similar length \eqref{eq:quantile-map}.  This prepares the features for the subsequent steps.

\subsection{Data joining, compression, and storing}
In this step, the commercial and experience datasets are loaded and left-joined using the primary key, namely the subscriber identifier. Next, the network data is then incorporated, with the subscriber's favorite site identifier acting as a secondary join key. The result is a large consolidated dataset---or the transformed design matrix $\mathbf{X}^{(Q)}$ as in \eqref{eq:transformed_matrix}---that is stored in columnar format within a PaaS object storage bucket.  This dataset is clearly no longer time based and has a uniform dimension.  The supertable is the augmentation of this design matrix by introducing the churn label as a supervisory signal.  In other words, $\mathbf{D} := [\mathbf{X}^{(Q)} \,\Vert\, \mathbf{y}]$, where the churn label $\mathbf{y}$ is based on the \textit{net} churner definition evaluated over the TP defined in Section~\ref{sec:introduction}.  Specifically, the label indicates whether a subscriber is expected to churn at any point during the TP.  The integration of the churn label requires careful alignment, not only on subscriber identity but also on the temporal relationship between historical data and the churn outcome as discussed earlier.  This dataset, aggregated over a \textit{window}, is then stored in a columnar format.  Owed to this module, the resulting learning task can therefore be framed as an example as:  Given subscribers' behaviors from July to December of last year, which subscribers are likely to churn during the subsequent January to March period?

\subsection{AI model training}
The supertable data is first cleansed ensuring no duplicate subscriber identifiers or subscribers with many missing value data points while preserving the class balance.
The training of the XGBoost model takes place next.  For the model training, a substantial portion of the supertable is divided into training and cross-validation sets, while the remaining data is reserved for testing and model scoring.  The performance measures are the recall and the F\textsubscript{1} score.  For model calibration, training is performed using stratified $K$-fold cross-validation combined with randomized hyperparameter search over a broad parameter space.  Stratification ensures that class balance is preserved across folds, enabling reliable hyperparameter selection and improved model performance. \review{The optimization objective is to find a threshold $\tau$ that maximizes the F\textsubscript{1} score on the training data.  Formally:
\begin{equation}
    \tau^\star \coloneqq \underset{\tau\in (0,1)}{\arg\,\max}\; F_1(\tau). 
    \label{eq:optimizer}
\end{equation}}%
The performance on the training data establishes our \textit{baseline} performance.%

\begin{figure*}[!t]
\centering
\resizebox{0.22\textwidth}{!}{\begin{tikzpicture}
\begin{axis}[
width=9cm, height=7cm,
legend columns=1,
legend cell align={left},
legend style={at={(0.76,0.28)},anchor=north east,fill opacity=0.6, draw opacity=1, font=\large, text opacity=1, draw=white!80!black, nodes={scale=0.6, transform shape}},
tick align=outside,
tick pos=left,
scale=0.5,
x grid style={white!69.0196078431373!black},
xlabel={Recall},
xmajorgrids,
xmin=0, xmax=1.03,
xtick={0,0.2,0.4,0.6,0.8,1.0},
ytick={0,0.2,0.4,0.6,0.8,1.0},
xminorgrids,
xtick style={color=black},
y grid style={white!69.0196078431373!black},
ylabel={Precision},
ymajorgrids,
ymin=0, ymax=1.03,
yminorgrids,
ytick style={color=black}
]

\addplot [color=blue, line width=1pt, mark=none]
table {%
0.0 1.0
0.1687 0.9981
0.3984 0.9943
0.4818 0.9906
0.5365 0.9868
0.5833 0.983
0.6146 0.9792
0.638 0.9755
0.6641 0.9717
0.6849 0.9679
0.7005 0.9642
0.7161 0.9604
0.7318 0.9566
0.7448 0.9528
0.7604 0.9491
0.7708 0.9453
0.7786 0.9415
0.7917 0.9377
0.7995 0.934
0.8073 0.9302
0.8177 0.9264
0.8229 0.9226
0.8307 0.9189
0.8359 0.9151
0.8411 0.9113
0.8516 0.9075
0.8568 0.9038
0.862 0.9
0.8672 0.8962
0.8724 0.8925
0.875 0.8887
0.8776 0.8849
0.8828 0.8811
0.888 0.8774
0.8932 0.8736
0.8958 0.8698
0.8984 0.866
0.9036 0.8623
0.9062 0.8585
0.9089 0.8547
0.9141 0.8509
0.9141 0.8472
0.9193 0.8434
0.9193 0.8396
0.9245 0.8358
0.9271 0.8321
0.9297 0.8283
0.9323 0.8245
0.9349 0.8208
0.9349 0.817
0.9375 0.8132
0.9401 0.8094
0.9427 0.8057
0.9453 0.8019
0.9479 0.7981
0.9505 0.7943
0.9505 0.7906
0.9505 0.7868
0.9531 0.783
0.9557 0.7792
0.9557 0.7755
0.9609 0.7717
0.9609 0.7679
0.9609 0.7642
0.9635 0.7604
0.9661 0.7566
0.9661 0.7528
0.9661 0.7491
0.9688 0.7453
0.9688 0.7415
0.9714 0.7377
0.9714 0.734
0.974 0.7302
0.974 0.7264
0.9766 0.7226
0.9766 0.7189
0.9766 0.7151
0.9792 0.7113
0.9792 0.7075
0.9818 0.7038
0.9818 0.7
0.9818 0.6962
0.9818 0.6925
0.9818 0.6887
0.9844 0.6849
0.9844 0.6811
0.9844 0.6774
0.9844 0.6736
0.987 0.6698
0.987 0.6585
0.9896 0.6472
0.9922 0.6434
0.9922 0.6245
0.9948 0.6208
0.9948
0.9948 0.6019
0.9974 0.5981
0.9974 0.5377
1.0 0.534
1.0 0.5302
1.0 0
};
\addlegendentry{PR (crossval)}
 
\draw[dashed, gray, thin] (axis cs:0.3221,0.447) -- (axis cs:0.3221,0.9961);
\draw[dashed, gray, thin] (axis cs:-0.047,0.9961) -- (axis cs:0.3221,0.9961);
 
\addplot[
    only marks,
    mark=*,
    mark size=2.3pt,
    color=red,
] table [x=recall, y=precision] {
recall precision
0.3221 0.9961
};
\addlegendentry{Best threshold $\tau^\star$}
 
\end{axis}
\end{tikzpicture}}%
\hfil
\resizebox{0.22\textwidth}{!}{\begin{tikzpicture}
\begin{axis}[
width=9cm, height=7cm,
legend columns=1,
legend cell align={left},
legend style={at={(0.76,0.28)},anchor=north east,fill opacity=0.6, draw opacity=1, font=\large, text opacity=1, draw=white!80!black, nodes={scale=0.6, transform shape}},
tick align=outside,
tick pos=left,
scale=0.5,
x grid style={white!69.0196078431373!black},
xlabel={Recall},
xmajorgrids,
xmin=0, xmax=1.03,
xtick={0,0.2,0.4,0.6,0.8,1.0},
ytick={0,0.2,0.4,0.6,0.8,1.0},
xminorgrids,
xtick style={color=black},
y grid style={white!69.0196078431373!black},
ylabel={Precision},
ymajorgrids,
ymin=0, ymax=1.03,
yminorgrids,
ytick style={color=black}
]

\addplot [color=blue, line width=1pt, mark=none]
table {%
0.0 1.0
0.0804 1.0
0.1583 0.9962
0.2412 0.9924
0.2965 0.9886
0.3442 0.9848
0.3819 0.9811
0.4121 0.9773
0.4397 0.9735
0.4573 0.9697
0.4749 0.9659
0.4874 0.9621
0.5 0.9583
0.5101 0.9545
0.5176 0.9508
0.5302 0.947
0.5402 0.9432
0.5503 0.9394
0.5578 0.9356
0.5704 0.9318
0.5804 0.928
0.5879 0.9242
0.598 0.9205
0.608 0.9167
0.6181 0.9129
0.6256 0.9091
0.6332 0.9053
0.6432 0.9015
0.6533 0.8977
0.6633 0.8939
0.6709 0.8902
0.6784 0.8864
0.6884 0.8826
0.6985 0.8788
0.706 0.875
0.7136 0.8712
0.7236 0.8674
0.7337 0.8636
0.7387 0.8598
0.7487 0.8561
0.7588 0.8523
0.7638 0.8485
0.7739 0.8447
0.7814 0.8409
0.7889 0.8371
0.794 0.8333
0.8015 0.8295
0.809 0.8258
0.8141 0.822
0.8216 0.8182
0.8266 0.8144
0.8317 0.8106
0.8392 0.8068
0.8467 0.803
0.8492 0.7992
0.8568 0.7955
0.8618 0.7917
0.8668 0.7879
0.8719 0.7841
0.8769 0.7803
0.8819 0.7765
0.8869 0.7727
0.8894 0.7689
0.8945 0.7652
0.897 0.7614
0.902 0.7576
0.9196 0.7424
0.9221 0.7386
0.9246 0.7348
0.9271 0.7311
0.9322 0.7273
0.9322 0.7235
0.9372 0.7197
0.9397 0.7159
0.9422 0.7121
0.9472 0.7083
0.9472 0.7045
0.9523 0.7008
0.9523 0.697
0.9548 0.6932
0.9573 0.6894
0.9598 0.6856
0.9623 0.6818
0.9623 0.678
0.9673 0.6742
0.9673 0.6705
0.9673 0.6667
0.9698 0.6629
0.9724 0.6591
0.9724 0.6553
0.9724 0.6515
0.9749 0.6477
0.9774 0.6439
0.9774 0.6402
0.9799 0.6364
0.9799 0.6326
0.9824 0.6288
0.9824 0.625
0.9824 0.6212
0.9849 0.6174
0.9849 0.6136
0.9874 0.6098
0.9874 0.6061
0.9874 0.6023
0.9874 0.5985
0.9874 0.5947
0.9874 0.5909
0.9874 0.5871
0.9899 0.5833
0.9899 0.5795
0.9899 0.5758
0.9899 0.572
0.9925 0.5682
0.9925 0.5644
0.9925 0.5606
0.9925 0.5568
0.9925 0.553
0.9925 0.5492
0.995 0.5265
0.9975 0.5227
1.0 0.4545
1.0 0
};
\addlegendentry{PR (inference)}
 
\draw[dashed, gray, thin] (axis cs:0.9119,0) -- (axis cs:0.9119,0.7493);
\draw[dashed, gray, thin] (axis cs:-0.05,0.7493) -- (axis cs:0.9119,0.7493);
 
\addplot[
    only marks,
    mark=*,
    mark size=2.3pt,
    color=red,
]
coordinates {(0.9119, 0.7493)};
\addlegendentry{Best threshold $\tau^\star$}
 
\end{axis}
\end{tikzpicture}}%
\hfil
\resizebox{0.22\textwidth}{!}{\begin{tikzpicture}
\begin{axis}[
width=9cm, height=7cm,
legend columns=1,
legend cell align={left},
legend style={at={(0.9,0.48)},anchor=north east,fill opacity=0.6, draw opacity=1, text opacity=1, draw=white!80!black, nodes={scale=0.6, transform shape}},
tick align=outside,
tick pos=left,
scale=0.5,
x grid style={white!69.0196078431373!black},
xlabel={Recall},
xmajorgrids,
xmin=0, xmax=1.03,
xtick={0,0.2,0.4,0.6,0.8,1.0},
ytick={0,0.2,0.4,0.6,0.8,1.0},
xminorgrids,
xtick style={color=black},
y grid style={white!69.0196078431373!black},
ylabel={Precision},
ymajorgrids,
ymin=0, ymax=1.03,
yminorgrids,
ytick style={color=black}
]

\addplot [color=blue, line width=1pt, mark=none]
table {%
0.0 1.0
0.0 1.0
0.0 0.9932
0.0 0.9864
0.0 0.9797
-0.0028 0.9729
-0.0028 0.9661
-0.0028 0.9593
-0.0028 0.9525
-0.0028 0.9458
0.0 0.939
-0.0028 0.9322
-0.0028 0.9254
0.0 0.9186
0.0 0.9119
0.0 0.9051
0.0 0.8983
0.0 0.8915
-0.0028 0.8847
-0.0028 0.878
-0.0028 0.8712
-0.0028 0.8644
-0.0028 0.8576
0.0 0.8508
0.0113 0.8847
0.0169 0.8644
0.0226 0.8475
0.0282 0.8271
0.0339 0.8169
0.0395 0.8068
0.0452 0.8
0.0508 0.7932
0.0565 0.7864
0.0621 0.7831
0.0678 0.7797
0.0734 0.7797
0.0791 0.7729
0.0847 0.7695
0.0904 0.7661
0.096 0.7593
0.1017 0.7525
0.1073 0.7424
0.113 0.7288
0.1186 0.7119
0.1243 0.6949
0.1299 0.6746
0.1356 0.6508
0.1412 0.6271
0.1469 0.6068
0.1525 0.5898
0.1582 0.5763
0.1638 0.5593
0.1695 0.5424
0.1751 0.5288
0.1808 0.5153
0.1864 0.5085
0.1921 0.5017
0.1977 0.4949
0.2034 0.4949
0.209 0.4915
0.2147 0.4915
0.2203 0.4915
0.226 0.4915
0.2316 0.4915
0.2373 0.4915
0.2429 0.4915
0.2486 0.4915
0.2542 0.4949
0.2599 0.4949
0.2655 0.4949
0.2712 0.4949
0.2768 0.4949
0.2825 0.4949
0.2881 0.4949
0.2938 0.4949
0.2994 0.4949
0.3051 0.4949
0.3107 0.4983
0.3164 0.4983
0.322 0.4983
0.3277 0.4983
0.3333 0.4983
0.339 0.4983
0.3446 0.4983
0.3503 0.4983
0.3559 0.4983
0.3616 0.4983
0.3672 0.4983
0.3729 0.4983
0.3785 0.4983
0.3842 0.4983
0.3898 0.4983
0.3955 0.4983
0.4011 0.4983
0.4068 0.4983
0.4124 0.4983
0.4181 0.4983
0.4237 0.4983
0.4294 0.4983
0.435 0.4983
0.4407 0.4983
0.4463 0.4983
0.452 0.4983
0.4576 0.4949
0.4633 0.4949
0.4689 0.4949
0.4746 0.4949
0.4802 0.4949
0.4859 0.4949
0.4915 0.4949
0.4972 0.4949
0.5028 0.4949
0.5085 0.4949
0.5141 0.4949
0.5198 0.4915
0.5254 0.4915
0.5311 0.4915
0.5367 0.4915
0.5424 0.4915
0.548 0.4915
0.5537 0.4915
0.5593 0.4881
0.565 0.4881
0.5706 0.4881
0.5763 0.4881
0.5819 0.4881
0.5876 0.4881
0.5932 0.4881
0.5989 0.4881
0.6045 0.4881
0.6102 0.4847
0.6158 0.4847
0.6215 0.4847
0.6271 0.4847
0.6328 0.4847
0.6384 0.4847
0.6441 0.4847
0.6497 0.4847
0.6554 0.4814
0.661 0.4814
0.6667 0.4814
0.6723 0.478
0.678 0.478
0.6836 0.4746
0.6893 0.4746
0.6949 0.4746
0.7006 0.4712
0.7062 0.4712
0.7119 0.4712
0.7175 0.4678
0.7232 0.4678
0.7288 0.4644
0.7345 0.4644
0.7401 0.4644
0.7627 0.4542
0.7684 0.4508
0.774 0.4475
0.7797 0.4475
0.7853 0.4441
0.791 0.4407
0.7966 0.4373
0.8023 0.4373
0.8079 0.4339
0.8136 0.4305
0.8192 0.4271
0.8249 0.4237
0.8305 0.4203
0.8362 0.4136
0.8418 0.4136
0.8475 0.4068
0.8531 0.4034
0.8588 0.4
0.8644 0.3932
0.8701 0.3898
0.8757 0.3864
0.8814 0.3797
0.887 0.3763
0.8927 0.3729
0.8983 0.3661
0.904 0.3593
0.9096 0.3525
0.9153 0.3458
0.9209 0.339
0.9266 0.3322
0.9322 0.3254
0.9379 0.3186
0.9435 0.3085
0.9492 0.2983
0.9548 0.2915
0.9605 0.2814
0.9661 0.2678
0.9718 0.2576
0.9774 0.2407
0.9831 0.2237
0.9887 0.2068
0.9944 0.1831
1 0 
};
\addlegendentry{PR (Case 3-1)}

\addplot [color=black, line width=1pt, mark=none]
table {%
0.0 1.0
-0.0027 0.9939
-0.0027 0.9878
-0.0027 0.9817
-0.0027 0.9756
-0.0027 0.9695
-0.0027 0.9634
-0.0027 0.9573
-0.0027 0.9511
0.0 0.945
0.0053 0.9969
0.0106 0.9969
0.016 0.9969
0.0213 0.9969
0.0266 0.9969
0.0319 0.9969
0.0372 0.9969
0.0426 0.9969
0.0479 0.9969
0.0532 0.9969
0.0585 0.9969
0.0638 0.9969
0.0691 0.9969
0.0745 0.9969
0.0798 0.9969
0.0851 0.9969
0.0904 0.9969
0.0957 0.9969
0.1011 0.9969
0.1064 0.9969
0.1117 0.9939
0.117 0.9939
0.1223 0.9939
0.1277 0.9939
0.133 0.9939
0.1383 0.9939
0.1436 0.9939
0.1489 0.9939
0.1543 0.9908
0.1596 0.9908
0.1649 0.9908
0.1702 0.9908
0.1755 0.9908
0.1809 0.9908
0.1862 0.9878
0.1915 0.9878
0.1968 0.9878
0.2021 0.9878
0.2074 0.9878
0.2128 0.9847
0.2181 0.9847
0.2234 0.9847
0.2287 0.9847
0.234 0.9847
0.2394 0.9817
0.2447 0.9786
0.25 0.9786
0.2553 0.9756
0.2606 0.9725
0.266 0.9725
0.2713 0.9695
0.2766 0.9664
0.2819 0.9664
0.2872 0.9634
0.2926 0.9603
0.2979 0.9573
0.3032 0.9542
0.3085 0.9511
0.3138 0.9511
0.3191 0.9481
0.3245 0.945
0.3298 0.942
0.3351 0.942
0.3404 0.9389
0.3457 0.9359
0.3511 0.9328
0.3564 0.9328
0.3617 0.9298
0.367 0.9267
0.3723 0.9237
0.3777 0.9237
0.383 0.9206
0.3883 0.9176
0.3936 0.9176
0.3989 0.9176
0.4043 0.9145
0.4096 0.9145
0.4149 0.9115
0.4202 0.9115
0.4255 0.9084
0.4309 0.9053
0.4362 0.9023
0.4415 0.9023
0.4468 0.8992
0.4521 0.8992
0.4574 0.8962
0.4628 0.8962
0.4681 0.8931
0.4734 0.8901
0.4787 0.8901
0.484 0.887
0.4894 0.884
0.4947 0.884
0.5 0.8809
0.5053 0.8809
0.5106 0.8809
0.516 0.8779
0.5213 0.8748
0.5266 0.8748
0.5319 0.8718
0.5372 0.8687
0.5426 0.8687
0.5479 0.8656
0.5532 0.8656
0.5585 0.8626
0.5638 0.8626
0.5691 0.8595
0.5745 0.8565
0.5798 0.8565
0.5851 0.8534
0.5904 0.8504
0.5957 0.8473
0.6011 0.8473
0.6064 0.8443
0.6117 0.8443
0.617 0.8412
0.6223 0.8382
0.6277 0.8351
0.633 0.8321
0.6383 0.829
0.6436 0.826
0.6489 0.826
0.6543 0.8198
0.6596 0.8168
0.6649 0.8137
0.6702 0.8107
0.6755 0.8076
0.6809 0.8046
0.6862 0.8015
0.6915 0.7954
0.6968 0.7954
0.7021 0.7893
0.7074 0.7863
0.7128 0.7832
0.7181 0.7771
0.7234 0.774
0.7287 0.771
0.734 0.7649
0.7394 0.7588
0.7447 0.7557
0.75 0.7527
0.7713 0.7282
0.7766 0.7221
0.7819 0.716
0.7872 0.7099
0.7926 0.7038
0.7979 0.6977
0.8032 0.6916
0.8085 0.6855
0.8138 0.6794
0.8191 0.6733
0.8245 0.6672
0.8298 0.658
0.8351 0.6519
0.8404 0.6427
0.8457 0.6336
0.8511 0.6275
0.8564 0.6183
0.8617 0.6092
0.867 0.6
0.8723 0.5908
0.8777 0.5847
0.883 0.5725
0.8883 0.5664
0.8936 0.5542
0.8989 0.542
0.9043 0.5359
0.9096 0.5237
0.9149 0.5115
0.9202 0.4992
0.9255 0.4901
0.9309 0.4748
0.9362 0.4626
0.9415 0.4473
0.9468 0.4351
0.9521 0.4198
0.9574 0.4015
0.9628 0.3863
0.9681 0.3649
0.9734 0.3466
0.9787 0.3221
0.984 0.2916
0.9894 0.255
0.9947 0.2061
1.0 0.1817
};
\addlegendentry{PR (Case 3-2)}

\draw[dashed, gray, thin] (axis cs:0.7514,0.08) -- (axis cs:0.7514,0.4576);
\draw[dashed, gray, thin] (axis cs:-0.04,0.4576) -- (axis cs:0.7514,0.4576);

\addplot[
    only marks,
    mark=*,
    mark size=2.3pt,
    color=red,
] table [x=recall, y=precision] {
recall precision
0.7514 0.4576
};
\addlegendentry{Best threshold (Case 3-1) $\tau^\star$}

\draw[dashed, gray, thin] (axis cs:0.762,0.13) -- (axis cs:0.762,0.7394);
\draw[dashed, gray, thin] (axis cs:-0.045,0.7394) -- (axis cs:0.762,0.7394);

\addplot[
    only marks,
    mark=+,
    mark size=2.3pt,
    color=red,
] table [x=recall, y=precision] {
recall precision
0.762 0.7394
};
\addlegendentry{Best threshold (Case 3-2) $\tau^\star$}

\end{axis}
\end{tikzpicture}}%
\hfil
\resizebox{0.22\textwidth}{!}{\begin{tikzpicture}
\begin{axis}[
width=9cm, height=7cm,
legend columns=1,
legend cell align={left},
legend style={at={(0.9,0.48)},anchor=north east,fill opacity=0.6, draw opacity=1, font=\small, text opacity=1, draw=white!80!black, nodes={scale=0.6, transform shape}},
tick align=outside,
tick pos=left,
scale=0.5,
x grid style={white!69.0196078431373!black},
xlabel={Recall},
xmajorgrids,
xmin=0, xmax=1.03,
xtick={0,0.2,0.4,0.6,0.8,1.0},
ytick={0,0.2,0.4,0.6,0.8,1.0},
xminorgrids,
xtick style={color=black},
y grid style={white!69.0196078431373!black},
ylabel={Precision},
ymajorgrids,
ymin=0, ymax=1.03,
yminorgrids,
ytick style={color=black}
]

\addplot [color=blue, line width=1pt, mark=none]
table {%
0 1
0.48 1
0.73 0.78
0.93 0.59
1 0.5
1 0
};
\addlegendentry{PR (Case 4-1)}

\addplot [color=black, line width=1pt, mark=none]
table {%
0 1
0.49 1
0.73 0.79
0.93 0.56
1 0.5
1 0
};
\addlegendentry{PR (Case 4-2)}
 
\draw[dashed, gray, thin] (axis cs:0.73,0.10) -- (axis cs:0.73,0.78);
\draw[dashed, gray, thin] (axis cs:0.00,0.78) -- (axis cs:0.73,0.78);
 
\addplot[
    only marks,
    mark=*,
    mark size=2.3pt,
    color=red,
]
coordinates {(0.73,0.78)};
\addlegendentry{Best threshold (Case 4-1) $\tau^\star$}

\addplot[
    only marks,
    mark=+,
    mark size=2.3pt,
    color=red,
]
 coordinates {(0.73,0.78)};
\addlegendentry{Best threshold (Case 4-2) $\tau^\star$}

\end{axis}
\end{tikzpicture}}%
\caption{Area under the precision-recall curve in the cases: Case 1 (training), Case 2 (inference on past unseen data), Cases 3-1 and 3-2 (inference over the next two months), and Cases 4-1 and 4-2 (inference over the next two months with truncation). \review{The marked point on each inference panel is the operating point given by the training-derived thresholds.}}
\label{fig:perf}
\end{figure*}

\begin{figure*}[!t]
\centering
\resizebox{0.22\textwidth}{!}{\begin{tikzpicture}
\begin{axis}[
    width=10cm, height=6.2cm,
    xlabel={Decile},
    ylabel={Lift over baseline},
    ybar,
    symbolic x coords={D1,D2,D3,D4,D5,D6,D7,D8,D9,D10},
    xtick=data,
    ymin=0, ymax=7,
    ytick={0,1,2,3,4,5,6},
    enlarge x limits=0.08,
    ymajorgrids=true,
    grid style={dashed, gray!40},
    tick align=outside,
    tick pos=left,
    bar width=12pt,
    axis on top
]

\addplot[fill={rgb,255:red,31;green,119;blue,180}, draw=none] coordinates {
(D1,2.13)
(D2,2.12)
(D3,2.03)
(D4,1.73)
(D5,1.16)
(D6,0.57)
(D7,0.25)
(D8,0.12)
(D9,0.07)
(D10,0.02)
};

\draw[red, dashed, thick] (rel axis cs:0,0.17241) -- (rel axis cs:1,0.17241);

\end{axis}
\end{tikzpicture}}%
\hfil
\resizebox{0.22\textwidth}{!}{\begin{tikzpicture}
\begin{axis}[
    width=10cm, height=6.2cm,
    xlabel={Decile},
    ylabel={Lift over baseline},
    ybar,
    symbolic x coords={D1,D2,D3,D4,D5,D6,D7,D8,D9,D10},
    xtick=data,
    ymin=0, ymax=5.8,
    ytick={0,1,2,3,4,5},
    enlarge x limits=0.08,
    ymajorgrids=true,
    grid style={dashed, gray!40},
    tick align=outside,
    tick pos=left,
    bar width=12pt,
    axis on top
]

\addplot[fill={rgb,255:red,31;green,119;blue,180}, draw=none] coordinates {
(D1,5.394)
(D2,2.913)
(D3,0.984)
(D4,0.433)
(D5,0.197)
(D6,0.118)
(D7,0.079)
(D8,0.039)
(D9,0.02)
(D10,0.01)
};

\draw[red, dashed, thick] (rel axis cs:0,0.17241) -- (rel axis cs:1,0.17241);

\end{axis}
\end{tikzpicture}}%
\hfil
\resizebox{0.22\textwidth}{!}{\begin{tikzpicture}
\begin{axis}[
    width=10cm, height=6.2cm,
    xlabel={Decile},
    ylabel={Lift over baseline},
    ybar,
    symbolic x coords={D1,D2,D3,D4,D5,D6,D7,D8,D9,D10},
    xtick=data,
    ymin=0, ymax=7,
    ytick={0,1,2,3,4,5,6},
    enlarge x limits=0.08,
    ymajorgrids=true,
    grid style={dashed, gray!40},
    tick align=outside,
    tick pos=left,
    bar width=12pt,
    axis on top
]

\addplot[fill={rgb,255:red,31;green,119;blue,180}, draw=none] coordinates {
(D1,4)
(D2,3.4)
(D3,1.5)
(D4,0.6)
(D5,0.28)
(D6,0.15)
(D7,0.06)
(D8,0.03)
(D9,0.02)
(D10,0.01)
};

\draw[red, dashed, thick] (rel axis cs:0,0.17241) -- (rel axis cs:1,0.17241);

\end{axis}
\end{tikzpicture}}%
\hfil
\resizebox{0.22\textwidth}{!}{\begin{tikzpicture}
\begin{axis}[
    width=10cm, height=6.2cm,
    xlabel={Decile},
    ylabel={Lift over baseline},
    ybar,
    symbolic x coords={D1,D2,D3,D4,D5,D6,D7,D8,D9,D10},
    xtick=data,
    ymin=0, ymax=7,
    ytick={0,1,2,3,4,5,6},
    enlarge x limits=0.08,
    ymajorgrids=true,
    grid style={dashed, gray!40},
    tick align=outside,
    tick pos=left,
    bar width=12pt,
    axis on top
]

\addplot[fill={rgb,255:red,31;green,119;blue,180}, draw=none] coordinates {
(D1,6.52)
(D2,2.52)
(D3,1.24)
(D4,0)
(D5,0)
(D6,0)
(D7,0)
(D8,0)
(D9,0)
(D10,0)
};

\draw[red, dashed, thick] (rel axis cs:0,0.17241) -- (rel axis cs:1,0.17241);

\end{axis}
\end{tikzpicture}}%
\caption{Decile lift plots for the cases: Case 1 (training), Case 2 (inference on past unseen data), Case 3-1 (inference over the next one month), and Case 4-1 (inference over the next month with truncation).}
\label{fig:decile_lift}
\end{figure*}

\subsection{Inferencing and insights} 
The remaining portion of the supertable after the training and cross-validation sets are dropped is used for the testing and model scoring.  Specifically, this data portion is used for evaluating the predictive capability of the model and for detecting overfitting or underfitting. Although the prediction task is described in forward-looking terms, all data involved in this evaluation thus far originates from historical records. In effect, the model is assessed on its ability to ``predict the past,'' since true churn outcomes cannot be observed for future periods at the time of evaluation.  Once the model demonstrates strong performance in predicting historical churn, the logical next step is to apply it to future churn prediction (i.e., in \textit{inferencing} mode), which is on the true holdout (or unseen) dataset. This is where the true value of a churn prediction model emerges: providing telcos with sufficient advance notice to proactively engage subscribers who are at high risk of churning.  

Insights generation and explainability are done through sorting the top features by XGBoost feature importance, running SHAP \cite{NIPS2017_8a20a862,shap}, identifying the impact the various features have on churn and retention through ``beeswarm plots''.  More discussion is in Section~\ref{sec:results}.%

\begin{table*}[!t]
\centering
\setlength\doublerulesep{0.5pt}
\caption{Examples of metadata mapping}
\vspace*{-0.5em}
\label{tab:meta}
\begin{tabular}{lll} 
\hhline{===}
Feature & English & Group \\
\hline
$\mathbf{x}_1$ & subscriber tenure (in days) & Commercial Data \\
$\mathbf{x}_2$ & revenue in the current month (in USDs) & Commercial Data \\
$\mathbf{x}_3$ & quality of experience (in \%) & Experience Data \\
$\mathbf{x}_4$ & number of radio resource control connected users & Network Data \\
$\mathbf{x}_5$ & subsidiary brand & Commercial Data \\
$\mathbf{x}_6$ & LTE block error rate downlink (in \%) & Network Data \\
$\mathbf{x}_7$ & GSM handover setup rate & Network Data \\
\hhline{===}
\end{tabular}
\end{table*}%

\vspace*{-0.5em}
\subsection{Mapping to TM Forum} Operationalizing subscriber value optimization within the TM Forum Autonomous Networks (AN) architecture outlined in Section~\ref{sec:autonomous} requires deploying the combined churn prediction, uplift analysis, and recommendation engine within the Business Operations Intelligence layer. Churn prediction estimates the likelihood of subscriber churn in the absence of an intervention, while uplift analysis estimates the incremental change in subscriber value that can be attributed to an intervention. Operating at this layer allows the system to analyze subscriber migration matrices, predict churn risk, quantify causal uplift, and prescribe optimized commercial interventions.

\vspace*{-0.5em}
\section{Results}\label{sec:results} 

To test the relevance and validity of this framework, we implement a pipeline automating the architecture on a commercial PaaS using over $3{,}000$ lines of source code.  The data, as outlined in the problem statement, has approximately $800$ learning features and require a storage bucket exceeding $250$~TBs.  The data cannot leave the country in which it resides.  \review{Table~\ref{tab:repro} summarizes the settings needed to reproduce the pipeline.} %

In this section, we look at four cases, each of the data collection duration of a six-month and the churn flag collected for them in the TP of three subsequent months.  We give these cases a numerical annotation for simplicity:

\begin{enumerate}
    \item Case 1: training data (past data).
    \item Case 2: unseen inference data (past data).
    \item Case 3: inference periods (unseen future data).  We use a suffix (i.e., Case 3-1 and 3-2) for the first and second prediction periods, separated by a month. 
    \item Case 4: inference periods with truncation heuristic.  We similarly define Case 4-1 and 4-2.
\end{enumerate}

\begin{table}[!t]
\centering
\setlength\doublerulesep{0.5pt}
\caption{Grouping and representation of the top-$10$ predictive features by dataset category}
\vspace*{-0.5em}
\label{tab:predictor}
\begin{tabular}{lll} 
\hhline{===}
Group & \% of top-$10$ & \% of all \\
\hline
Commercial data & $90\%$ & $30\%$ \\
Experience data & $10$\% & $5\%$ \\
Network data & $0$\% & $65\%$ \\
\hhline{===}
\end{tabular}
\end{table}%

\begin{table*}[!t]
\centering
\setlength\doublerulesep{0.5pt}
\caption{Model performance and maintenance measures}
\vspace*{-0.5em}
\label{tab:performance}
\begin{tabular}{lcccccc} 
\hhline{=======}
& Case 1 & Case 2 & Case 3-1 & Case 3-2 & Case 4-1 & Case 4-2 \\
\cline{2-7}
F\textsubscript{1} & $0.8800$ & $0.8222$ & 0.7519 & $0.5664$ & $0.7591$ & $0.7542$ \\
Recall & $0.9110$ & $0.9118$ & 0.7627 & $0.7495$ & $0.7342$ & $0.7279$ \\
Features with PSI above threshold $0.45$ & - & $0.000\%$ & $0.2532\%$ & $0.253\%$& $0.253\%$ & $0.000\%$\\ 
Features with null ratio above threshold $0.02$ & - & $0.633\%$ & $0.000\%$ & $0.000\%$& $0.000\%$ & $0.759\%$ \\
\hhline{=======}
\end{tabular}
\end{table*}%

The target performance for our problem statement is an inference  F\textsubscript{1} score of $0.75$ with a recall of at least $0.70$ across all the network and a maximum reuse longevity. \review{In implementing the transform in \eqref{eq:quantile-map}, we use $(q_1, q_2, q_3, q_4, q_5) \coloneqq (0, 25, 50, 75, 100)$ and thus $Q = 5$ for a five-fold growth in the number of columns in the design matrix as a result.}

\subsection{Classification performance}
We start with the PR curves, which plot precision against recall across the full range of classification thresholds. In the churn context, precision is the fraction of subscribers flagged as churners who actually churned (i.e., true positives over predicted positives), while recall is the fraction of actual churners correctly identified (i.e., true positives over condition positives). We favor PR curves over receiver operating characteristic (ROC) curves because the churn problem is class-imbalanced.  Moreover, this curve also allows us to find the decision point below which a prediction is assigned to class zero (non-churn), as a trade-off between precision and recall: lowering the threshold increases recall (capturing more true churners) at the cost of precision, as a larger share of retained subscribers are misclassified as churners, while raising it does the reverse.  

As a secondary measure of the classification performance, we study the decile lift as shown in Fig.~\ref{fig:decile_lift}.  A decile lift chart shows how well the model ranks positive cases by probability.  It breaks predictions into 10 equal-sized groups (deciles) based on predicted probabilities, then computes the lift in positive rate vs. the average (i.e., the ratio).  Across all four cases, the lift is monotonically decreasing from D1 to roughly D3-D5, confirming that the model correctly rank-orders subscribers by churn propensity: the highest-scored decile is consistently the most enriched for actual churners, and lift decays toward the baseline (dashed red line at $1.0$) as predicted probability falls.  The sharper front-loading in Case 3-1 (i.e., the first inference period) suggests the model separates the highest-risk subscribers even more cleanly on unseen data than during training, though at the cost of discriminating well within the middle deciles, where lift quickly approaches 1 and provides little ranking information beyond the top few groups.  
This is how the truncation heuristic exploits this situation: rather than removing entire deciles, it removes the small subset of subscribers within them whose churn probability falls in the low-confidence region defined in Section~\ref{sec:framework}. As detailed later in this section, this affects only approximately $3\%$ of the subscriber base.  This group of subscribers is concentrated among the ambiguous, middle-decile cases that were diluting the ranking signal. In Case 4-1, the ten deciles are recomputed over this truncated population, yielding a modestly sharper and more monotonic lift curve through the middle deciles rather than the flatter behavior observed in Case 3-1.

\subsection{Top predictors}
\begin{figure}[!t]
\centering
\includegraphics[width=0.4\textwidth]{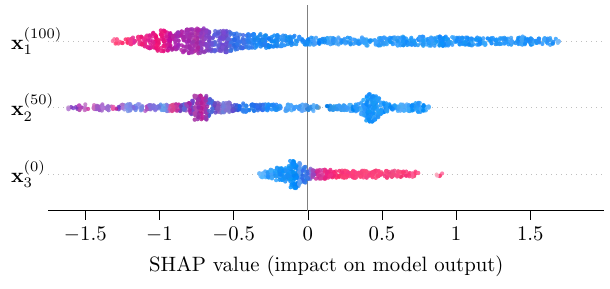}
\caption{SHAP showing select top predictors and their impact on churn. \review{The horizontal axis is the SHAP value on the model's log-odds (margin) scale.}}
\label{fig:shapley_output}
\end{figure}

Although model training incorporates commercial, experience, and network datasets, their contributions to churn prediction are not equal. In Fig.~\ref{fig:shapley_output}, we see how SHAP shows select top predictors belong to the commercial and experience dataset categories. This is applied across all the learning features.  The top ten predictors in our problem are dominated by the commercial dataset category despite it only representing a third of the learning features.  Conversely, the top ten predictors do not include any network data despite network data being the most represented in terms of the number of learning features.  Table~\ref{tab:predictor} shows more details.  \review{In our data and feature set, aggregated RAN counters carried less predictive signal than commercial and experience indicators, despite their greater representation.  The QoE index, which is derived from per-subscriber network measurements, ranked second in importance.}%

\begin{figure}[!t]
\centering
\resizebox{0.45\textwidth}{!}{\definecolor{churnorange}{HTML}{F4A76B}
\definecolor{churnblue}{HTML}{7FA8D9}
\definecolor{lineorange}{HTML}{E8791E}
\definecolor{lineblue}{HTML}{2C6FBB}

\begin{tikzpicture}
\begin{axis}[
  width=17cm, height=7.8cm,
  xmin=0, xmax=1, ymin=0, ymax=0.62,
  axis x line=bottom, axis y line=left,
  xtick={0,0.2,0.4,0.6,0.8,1.0},
  ytick={0,0.5},
  xticklabels={0.0,0.2,0.4,0.6,0.8,1.0},
  yticklabels={0.0,0.5},
  tick label style={font=\large\sffamily},
  axis line style={line width=1pt},
  clip=false,
]

\addplot[draw=none, fill=churnorange, fill opacity=0.55, domain=0:1, samples=200]
      {0.55*exp(-9*x)+0.005} \closedcycle;
\addplot[draw=none, fill=churnblue, fill opacity=0.55, domain=0:1, samples=200]
      {0.005 + 0.19/(1+exp(-30*(x-0.92)))} \closedcycle;

\addplot[lineorange, thick, domain=0:1, samples=200] {0.55*exp(-9*x)+0.005};
\addplot[lineblue, thick, domain=0:1, samples=200] {0.005 + 0.19/(1+exp(-30*(x-0.92)))};

\node[lineorange, font=\large\sffamily] at (axis cs:0.17,0.31) {$p(y=0\mid \mathbf{X})$};
\node[lineblue, font=\large\sffamily] at (axis cs:0.84,0.16) {$p(y=1\mid \mathbf{X})$};

\draw[black, line width=1pt] (axis cs:0.43,0.12) -- (axis cs:0.43,-0.20);
\draw[black, line width=1pt] (axis cs:0.50,0) -- (axis cs:0.50,0.42);
\draw[black, line width=1pt] (axis cs:0.57,0.12) -- (axis cs:0.57,-0.20);
\draw[lineblue, dashed, line width=0.9pt] (axis cs:0.02,0) -- (axis cs:0.02,-0.38);
\draw[lineblue, dashed, line width=0.9pt] (axis cs:0.98,0) -- (axis cs:0.98,-0.38);

\node[font=\bfseries\large] at (axis cs:0.50,0.53) {$F_1 \ge 0.69$};
\node[font=\Large] at (axis cs:0.50,0.47) {(0\% of subscribers dropped)};

\draw[black, {Latex[length=1.8mm]}-{Latex[length=1.8mm]}, line width=0.9pt]
      (axis cs:0.43,-0.20) -- (axis cs:0.57,-0.20);
\node[anchor=north, font=\Large] at (axis cs:0.50,-0.22)
      {$F_1 \ge 0.75$ \; (3\% of subscribers dropped)};

\draw[lineblue, {Latex[length=2mm]}-{Latex[length=2mm]}, line width=0.9pt]
      (axis cs:0.02,-0.34) -- (axis cs:0.98,-0.34);
\node[anchor=north, font=\Large] at (axis cs:0.50,-0.35)
      {$F_1 = 1.00$ \; (trivial: 98\% of subscribers dropped)};

\draw[black, fill=white] (axis cs:0.687,0.60) rectangle (axis cs:0.953,0.47);
\draw[fill=churnblue] (axis cs:0.70,0.565) rectangle (axis cs:0.745,0.585);
\node[anchor=west, font=\large] at (axis cs:0.755,0.575) {flag\_churn = $1$};
\draw[fill=churnorange] (axis cs:0.70,0.495) rectangle (axis cs:0.745,0.515);
\node[anchor=west, font=\large] at (axis cs:0.755,0.505) {flag\_churn = $0$};

\end{axis}
\end{tikzpicture}}%
\caption{\review{Truncation of low-confidence subscribers.  Subscribers are ranked by the probability assigned to their true class, and the lowest-ranked are dropped.  $F_1$ on the retained subscribers is shown against the fraction dropped: $F_1\approx0.69$ with $0\%$ dropped, $F_1\geq0.75$ with $3\%$ dropped, and $F_1=1.00$ with $98\%$ dropped.  The selection uses the true labels.}}%
\label{fig:truncation}
\end{figure}%

\begin{figure}[!t]
\centering
\resizebox{0.21\textwidth}{!}{\begin{tikzpicture}
\begin{axis}[
width=9cm, height=7cm,
legend columns=1,
legend cell align={left},
legend style={at={(0.76,0.28)},anchor=north east,fill opacity=0.6, draw opacity=1, font=\large, text opacity=1, draw=white!80!black, nodes={scale=0.6, transform shape}},
tick align=outside,
tick pos=left,
scale=0.5,
x grid style={white!69.0196078431373!black},
xlabel={Recall},
xmajorgrids,
xmin=0, xmax=1.03,
xtick={0,0.2,0.4,0.6,0.8,1.0},
ytick={0,0.2,0.4,0.6,0.8,1.0},
xminorgrids,
xtick style={color=black},
y grid style={white!69.0196078431373!black},
ylabel={Precision},
ymajorgrids,
ymin=0, ymax=1.03,
yminorgrids,
ytick style={color=black}
]

\addplot [color=blue, line width=1pt, mark=none]
table {%
0.0 0.9143
0.0 0.9107
0.0 0.9071
0.0 0.9036
0.0 0.9
0.0 0.8964
0.0 0.8929
0.0 0.8893
0.0 0.8857
0.0 0.8821
0.0 0.8786
0.0 0.875
0.0 0.8714
0.0 0.8679
0.0 0.8643
0.0 0.8607
0.0 0.8571
0.0 0.8536
0.0 0.85
0.0058 0.8482
0.0117 0.8429
0.0175 0.8304
0.0234 0.8232
0.0292 0.8179
0.0351 0.8143
0.0409 0.8089
0.0468 0.8036
0.0526 0.7982
0.0585 0.7946
0.0643 0.7911
0.0702 0.7875
0.076 0.7839
0.0819 0.7821
0.0877 0.7821
0.0936 0.7786
0.0994 0.7768
0.1053 0.7732
0.1111 0.7696
0.117 0.7679
0.1228 0.7661
0.1287 0.7625
0.1345 0.7607
0.1404 0.7571
0.1462 0.7554
0.152 0.7518
0.1579 0.75
0.1637 0.7482
0.1696 0.7446
0.1754 0.7429
0.1813 0.7411
0.1871 0.7375
0.193 0.7375
0.1988 0.7375
0.2047 0.7357
0.2105 0.7357
0.2164 0.7339
0.2222 0.7321
0.2281 0.7304
0.2339 0.7304
0.2398 0.7268
0.2456 0.725
0.2515 0.7232
0.2573 0.7196
0.2632 0.7179
0.269 0.7143
0.2749 0.7125
0.2807 0.7089
0.2865 0.7071
0.2924 0.7054
0.2982 0.7036
0.3041 0.7018
0.3099 0.7
0.3158 0.6982
0.3216 0.6964
0.3275 0.6946
0.3333 0.6911
0.3392 0.6875
0.345 0.6857
0.3509 0.6839
0.3567 0.6804
0.3626 0.6786
0.3684 0.6768
0.3743 0.675
0.3801 0.6732
0.386 0.6696
0.3918 0.6661
0.3977 0.6643
0.4035 0.6625
0.4094 0.6589
0.4152 0.6554
0.4211 0.6536
0.4269 0.6518
0.4327 0.6482
0.4386 0.6446
0.4444 0.6429
0.4503 0.6411
0.4561 0.6375
0.462 0.6357
0.4678 0.6339
0.4737 0.6304
0.4795 0.6286
0.4854 0.625
0.4912 0.6232
0.4971 0.6196
0.5029 0.6161
0.5088 0.6143
0.5146 0.6125
0.5205 0.6107
0.5263 0.6089
0.5322 0.6054
0.538 0.6036
0.5439 0.6018
0.5497 0.5982
0.5556 0.5964
0.5614 0.5946
0.5673 0.5929
0.5731 0.5911
0.5789 0.5893
0.5848 0.5875
0.5906 0.5839
0.5965 0.5821
0.6023 0.5804
0.6082 0.5786
0.614 0.5768
0.6199 0.5732
0.6257 0.5714
0.6316 0.5696
0.6374 0.5679
0.6433 0.5661
0.6491 0.5643
0.655 0.5607
0.6608 0.5589
0.6667 0.5571
0.6725 0.5554
0.6784 0.5536
0.6842 0.55
0.6901 0.5482
0.6959 0.5482
0.7018 0.5464
0.7076 0.5429
0.7135 0.5411
0.7193 0.5411
0.7251 0.5393
0.731 0.5375
0.7368 0.5357
0.7427 0.5339
0.7485 0.5321
0.7544 0.5304
0.7602 0.5268
0.7661 0.525
0.7719 0.525
0.7778 0.5232
0.7836 0.5214
0.7895 0.5196
0.7953 0.5179
0.8012 0.5161
0.807 0.5143
0.8129 0.5125
0.8187 0.5107
0.8246 0.5089
0.8304 0.5054
0.8363 0.5036
0.8421 0.5018
0.848 0.5018
0.8538 0.5
0.8596 0.4982
0.8655 0.4964
0.8713 0.4946
0.8772 0.4929
0.883 0.4929
0.8889 0.4911
0.8947 0.4875
0.9006 0.4857
0.9064 0.4857
0.9123 0.4839
0.9181 0.4821
0.924 0.4804
0.9298 0.4786
0.9357 0.4768
0.9415 0.475
0.9474 0.4732
0.9532 0.4714
0.9591 0.4714
0.9649 0.4696
0.9708 0.4679
0.9766 0.4661
0.9825 0.4643
1.0 0.4607
};
\addlegendentry{PR (crossval)}
 
\draw[dashed, gray, thin] (axis cs:1.0,0.44) -- (axis cs:1.0,0.935);
\draw[dashed, gray, thin] (axis cs:-0.04,0.4607) -- (axis cs:1.04,0.4607);

\addplot[
    only marks,
    mark=*,
    mark size=2.3pt,
    color=red,
] table [x=recall, y=precision] {
recall precision
1.0 0.4607
};
\addlegendentry{Best threshold $\tau^\star$}
 
\end{axis}
\end{tikzpicture}}%
\hfil
\resizebox{0.26\textwidth}{!}{\begin{tikzpicture}
\begin{axis}[
    width=10cm, height=6.2cm,
    xlabel={Decile},
    ylabel={Lift over baseline},
    ybar,
    symbolic x coords={D1,D2,D3,D4,D5,D6,D7,D8,D9,D10},
    xtick=data,
    ymin=0, ymax=7,
    ytick={0,1,2,3,4,5,6},
    enlarge x limits=0.08,
    ymajorgrids=true,
    grid style={dashed, gray!40},
    tick align=outside,
    tick pos=left,
    bar width=12pt,
    axis on top
]

\addplot[fill={rgb,255:red,31;green,119;blue,180}, draw=none] coordinates {
(D1,1.62)
(D2,1.45)
(D3,1.20)
(D4,1.04)
(D5,0.95)
(D6,0.87)
(D7,0.83)
(D8,0.76)
(D9,0.7)
(D10,0.68)
};

\draw[red, dashed, thick] (rel axis cs:0,0.17241) -- (rel axis cs:1,0.17241);

\end{axis}
\end{tikzpicture}}%
\caption{Ablation effect: Dropping commercial and experience data and keeping \review{all network data (RAN KPIs, core KPIs, and MDT)} shows a heavily biased classifier with poor performance.}
\label{fig:ran_only}
\end{figure}

Features derived from commercial datasets exhibit the strongest predictive power even though they are not the most represented in the learning features. Examples of these features are the subscriber tenure and recent account payment subscriber behavior (i.e., the telco revenue).

Experience-related features rank second in importance despite them ranking third in representation. Measures such as perceived voice quality, video quality, and buffering delays (e.g., in games and streaming) can strongly influence churn, especially when service degradation occurs repeatedly at locations or times that are most critical to the subscriber.

A notable observation from the analysis is that RAN busy-hour performance counters---which are by design aggregated at the cell site level and not the subscriber \cite{9500204,3gpp_ts_32425_rel16,3gpp_ts_28552_rel16}---exhibit limited predictive power for subscriber churn \review{in our data and feature set (busy-hour counters aggregated at each subscriber's favorite site, plus sparse MDT data)}.  This is true even though network datasets have the largest representation in the complete dataset.  This finding does not imply that network quality is irrelevant to churn prediction; rather, it highlights a measurement mismatch between network-centric KPIs and subscriber-perceived experience: Aggregated counters fail to capture spatial (i.e., cell center vs edge performance), temporal (e.g., peak vs off-peak), and device-level  capability or variability experienced by individual subscribers.  Moreover, MDT is sparse in nature due to subscribers behavior and mobility patterns.  This sparsity can lead to one of two scenarios: (1) the null ratio in these learning features is too high or (2) the variance as explained by them would be too small.  The former would cause these learning features to be dropped while the latter would cause them not to show among the top features by importance.   As a result of the aforementioned, we expect that models trained primarily on RAN features to perform poorly. To validate our claim, we show Fig.~\ref{fig:ran_only}.
\begin{figure*}[!t]
\centering
\resizebox{0.6\textwidth}{!}{

\begin{tikzpicture}[
    >=Stealth,
    thick,
    state/.style={circle, draw, minimum size=1cm},
    every node/.style={font=\small}
]

\node[state] (preamble)   at (0,0)    {};
\node[state] (mag_top)    at (3,0.9)  {};
\node[state] (mag_bot)    at (3,-0.9) {};
\node[state] (feature)    at (6,0)    {};
\node[state] (dir_top)    at (9,0.9)  {};
\node[state] (dir_bot)    at (9,-0.9) {};
\node[state] (conclusion) at (12,0)   {};
\node[state, double, double distance=2pt] (accept) at (14,0) {};

\draw[->] (preamble) -- node[above] {$p(a_0)$}     (mag_top);
\draw[->] (preamble) -- node[below] {$1-p(a_0)$}   (mag_bot);
\draw[->] (mag_top)  -- (feature);
\draw[->] (mag_bot)  -- (feature);
\draw[->] (feature)  -- node[above] {$p(a_1)$}     (dir_top);
\draw[->] (feature)  -- node[below] {$1-p(a_1)$}   (dir_bot);
\draw[->] (dir_top)  -- (conclusion);
\draw[->] (dir_bot)  -- (conclusion);
\draw[->] (conclusion) -- (accept);

\node[align=center, below=4pt of preamble]   {preamble};
\node[align=center, below=4pt of mag_bot]    {magnitude of\\ dominance};
\node[align=center, below=4pt of feature]    {feature name\\ and body of insight};
\node[align=center, below=4pt of dir_bot]    {direction\\ of relationship};
\node[align=center, below=4pt of conclusion] {conclusion};

\end{tikzpicture}}%
\caption{State diagram for English language insight generation.}
\label{fig:state_diag}
\end{figure*}%

\begin{figure*}[!t]
\centering
\resizebox{0.75\textwidth}{!}{\definecolor{navy}{HTML}{1F3864}
\definecolor{redline}{HTML}{C0504D}

\begin{tikzpicture}

\begin{axis}[
  name=leftaxis,
  scale only axis,
  width=20cm, height=6cm,
  xmin=1, xmax=36, ymin=0, ymax=120,
  xtick={1,...,36},
  xticklabel style={font=\small},
  ytick={0,20,40,60,80,100,120},
  yticklabel={\pgfmathprintnumber{\tick}\%},
  axis y line*=left,
  axis x line*=bottom,
  xlabel={Value segments (high to low)},
  ylabel={Cumulative subscriber distribution [\%]},
  label style={font=\small},
  tick label style={font=\small, navy},
  ylabel style={navy},
  grid=both,
  major grid style={line width=0.2pt, draw=black!15},
  axis line style={navy},
  tick style={navy},
]
\addplot[navy, line width=2, mark=diamond]
table{
1	0.083174784
2	0.490184526
3	1.580046973
4	14.79729435
5	14.81175678
6	14.89335776
7	15.07659448
8	17.94183992
9	17.99598125
10	18.08459059
11	18.41578724
12	21.29453559
13	23.635947
14	35.07431855
15	37.21212784
16	52.54272171
17	52.97373232
18	55.21182397
19	55.61758083
20	59.03712331
21	59.90296961
22	62.27901273
23	64.06556307
24	69.38655141
25	73.83979872
26	85.93798325
27	86.30135772
28	87.28797011
29	88.25659806
30	90.73435471
31	90.81749384
32	91.05816698
33	94.45277988
34	97.63565584
35	98.90140628
36	100

};
\end{axis}

\begin{axis}[
  at={(leftaxis.south west)}, anchor=south west,
  scale only axis,
  width=20cm, height=6cm,
  xmin=1, xmax=36, ymin=0, ymax=9,
  hide x axis,
  axis y line*=right,
  ylabel={Cumulative churn [\%]},
  ytick={0,2,4,6,8,10},
  yticklabel={\pgfmathprintnumber{\tick}\%},
  label style={font=\small},
  tick label style={font=\small, redline!80!black},
  ylabel style={redline!80!black},
  axis line style={redline!80!black},
  tick style={redline!80!black},
]
\addplot[redline, line width=2, mark=square*] 
table {
 1	0.003513902
2	0.012514225
3	0.031570015
4	0.276693883
5	0.278191506
6	0.283533675
7	0.292281153
8	0.431498466
9	0.450583429
10	0.461655139
11	0.509739523
12	0.762516827
13	0.811823296
14	1.051887016
15	1.099182068
16	1.434814857
17	1.458502903
18	1.587849522
19	1.611441941
20	1.810935323
21	2.093592143
22	2.451168993
23	3.26710751
24	4.212948581
25	4.357508064
26	4.675568099
27	4.692213753
28	4.727409505
29	4.802658559
30	4.970207521
31	4.979566041
32	5.000210218
33	6.577663905
34	7.403327244
35	8.293119303
36	8.794876367
};



\end{axis}

\node[fill=navy, minimum width=0.5cm, minimum height=0.13cm, inner sep=0pt]
      (sw1) at ($(leftaxis.south)+(-3.7,-1.2)$) {};
\node[anchor=west, font=\small] at (sw1.east) {\% cumulative base};

\node[fill=redline, minimum width=0.5cm, minimum height=0.13cm, inner sep=0pt]
      (sw2) at ($(leftaxis.south)+(-0.1,-1.2)$) {};
\node[anchor=west, font=\small] at (sw2.east) {\% cumulative churn};

\end{tikzpicture}}%
\caption{Churn rate increases as value contribution declines.}
\label{fig:val_contr}
\end{figure*}

\begin{figure*}[!t]
\centering
\resizebox{0.85\textwidth}{!}{\definecolor{navy}{HTML}{1F3864}
\definecolor{orangeline}{HTML}{F2A93B}

\begin{tikzpicture}

\begin{axis}[
  name=leftaxis,
  scale only axis,
  width=20cm, height=6cm,
  xmin=1, xmax=36, ymin=0, ymax=120,
  xtick={1,...,36},
  xticklabel style={font=\small},
  ytick={0,20,40,60,80,100,120},
  yticklabel={\pgfmathprintnumber{\tick}\%},
  axis y line*=left,
  axis x line*=bottom,
  xlabel={Value segments (high to low)},
  ylabel={Cumulative subscriber distribution [\%]},
  label style={font=\small},
  tick label style={font=\small, navy},
  ylabel style={navy},
  grid=both,
  major grid style={line width=0.2pt, draw=black!15},
  axis line style={navy},
  tick style={navy},
]
\addplot[navy, line width=2,mark=diamond]
table{
1	0.083174784
2	0.490184526
3	1.580046973
4	14.79729435
5	14.81175678
6	14.89335776
7	15.07659448
8	17.94183992
9	17.99598125
10	18.08459059
11	18.41578724
12	21.29453559
13	23.635947
14	35.07431855
15	37.21212784
16	52.54272171
17	52.97373232
18	55.21182397
19	55.61758083
20	59.03712331
21	59.90296961
22	62.27901273
23	64.06556307
24	69.38655141
25	73.83979872
26	85.93798325
27	86.30135772
28	87.28797011
29	88.25659806
30	90.73435471
31	90.81749384
32	91.05816698
33	94.45277988
34	97.63565584
35	98.90140628
36	100
};
\end{axis}

\begin{axis}[
  at={(leftaxis.south west)}, anchor=south west,
  scale only axis,
  width=20cm, height=6cm,
  xmin=1, xmax=36, ymin=-8, ymax=6,
  hide x axis,
  axis y line*=right,
  ylabel={Cumulative value leakage [\%]},
  ytick={-8,-6,-4,-2,0,2,4,6},
  yticklabel={\pgfmathprintnumber{\tick}\%},
  label style={font=\small},
  tick label style={font=\small, orangeline!80!black},
  ylabel style={orangeline!80!black},
  axis line style={orangeline!80!black},
  tick style={orangeline!80!black},
]
\addplot[orangeline, line width=2, mark=square*] 
table {
1	0.046329053
2	0.251389896
3	0.663905509
4	1.292589332
5	1.29897854
6	1.341702926
7	1.37976588
8	1.203775772
9	1.217886489
10	1.255082314
11	1.357332336
12	1.112380272
13	2.07404525
14	3.765525719
15	4.016422891
16	0.56051282
17	0.665442031
18	0.779711933
19	0.756255662
20	-0.372390996
21	-0.180165298
22	-0.110496672
23	0.03468682
24	-1.304815054
25	-0.714745258
26	-3.055472687
27	-3.20093982
28	-3.852519116
29	-3.915296085
30	-4.715091397
31	-4.753992307
32	-4.918923671
33	-4.644119645
34	-5.312850286
35	-5.465787703
36	-5.884162521
};



\end{axis}

\node[fill=navy, minimum width=0.5cm, minimum height=0.13cm, inner sep=0pt]
      (sw1) at ($(leftaxis.south)+(-3.7,-1.2)$) {};
\node[anchor=west, font=\small] at (sw1.east) {\% cumulative base};

\node[fill=orangeline, minimum width=0.5cm, minimum height=0.13cm, inner sep=0pt]
      (sw2) at ($(leftaxis.south)+(-0.1,-1.2)$) {};
\node[anchor=west, font=\small] at (sw2.east) {\% cumulative net leakage};

\end{tikzpicture}}%
\caption{Observed value leakages are the highest for high-value segments. \review{This pattern is descriptive and does not establish that an intervention would prevent churn.}}
\label{fig:val_leakage}
\end{figure*}

\subsection{Ablation effect}
To assess the impact of dropping \review{the commercial and experience data and keeping all network data (RAN KPIs, core KPIs, and MDT)} we show the plots in Fig.~\ref{fig:ran_only}.  Here we observe a PR curve declining almost linearly from a precision of $0.85$ at low recall down to $0.45$ at a perfect recall with no plateau or elbow. In other words, there is no optimal or ``sweet spot'' point where you get both good precision and good recall.

It is worth highlighting where $\tau^\star$ sits: at a recall of $1$ and a precision of $0.45$, where every subscriber is classified subscriber as churner, we have a diagnostic of poor class separation, since F\textsubscript{1} is being maximized by recall alone.  When studying the decile lift, we observe that only D1 shows a meaningful lift close to twice the baseline.  The rest of the deciles oscillate or flatten around the baseline, which is unlike a well-performing model that would show a smooth and monotonically decreasing lift.

Taken together, these results indicate that \review{in our data and feature set, the network features alone carry limited discriminative information for separating the subscriber population.}

\subsection{Windowing effect}
In Table~\ref{tab:performance}, we show the recall and F\textsubscript{1} scores using the true churn data, which for inference purposes is obtained after the predictions are made.

To study the model longevity, we observe the degradation in the performance measures over three inference periods---each spanning a duration of six months---defined using sliding windows that shift forward by one month in Table~\ref{tab:performance}.  The degradation is expected due to the age of information effect as discussed earlier.  Some increase in the ratio of features suffering from data drift is reported even with no features having null ratios higher than that of the training data.  The model has a time horizon of $2$ window slidings (or an inference of up to $5$ months), after which the F\textsubscript{1} score falls below the target performance thresholds stated earlier in this section.  \review{The model met the performance targets for up to two windows.  In the second window, precision fell sharply at the frozen threshold while recall was nearly unchanged; the PSI and null-ratio indicators do not explain this drop.}%

\subsection{Truncation effect}
\review{We cover the truncation diagnostic in Section~\ref{sec:framework}.  As shown in Fig.~\ref{fig:truncation}, it sets aside approximately $3\%$ of the subscribers, selected using the true labels, and $F_1$ on the retained subscribers meets the target in Table~\ref{tab:performance}.  This is an upper-bound diagnostic of the model on the cases it ranks confidently and is not comparable to $F_1$ on all subscribers.  The $F_1 = 1.00$ point at $98\%$ dropped (i.e., $2\%$ coverage) is the trivial limiting case and not a result.  Coverage is $97.24\%$ in Case 4-1 and $96.78\%$ in Case 4-2.  The PR curve outlining this effect is shown in Fig.~\ref{fig:perf} for Cases 3 and 4.  Table~\ref{tab:truncation} shows the percentages, which are shares of predicted labels under different cutoffs and subscriber sets and cannot be read as an effect of removal alone.}

\begin{table}[!t]
\centering
\setlength\doublerulesep{0.5pt}
\caption{\review{Truncation effect\textsuperscript{\textasteriskcentered}.  Percentages are shares of predicted labels over all subscribers scored in each case.}}
\vspace*{-0.5em}
\label{tab:truncation}
\begin{threeparttable}
\begin{tabular}{lccc} 
\hhline{====}
& $y = 0$ & $y = 1$ & $\tilde y = \mathtt{NaN}$ \\
\cline{2-4}
Case 3-1 (before truncation) & $89.40\%$ & $10.60\%$ & -\\
Case 4-1 (after truncation) & $82.95\%$ & $14.29\%$ & $2.76\%$ \\
\hline
Case 3-2 (before truncation) & $88.94\%$ & $11.06\%$ &  -\\
Case 4-2 (after truncation) & $84.01\%$ & $12.77\%$ & $3.22\%$ \\

\hhline{====}
\end{tabular}
\begin{tablenotes}\footnotesize
\item[\textasteriskcentered]{\review{ The ``before'' row is scored on all subscribers and the ``after'' row on the subscribers retained by truncation; the ``before'' row is classified at the optimal point of the precision--recall curve of its own set and the ``after'' row at the training-derived $\tau^\star$, so the cutoffs differ.  
About $3\%$ of subscribers are dropped and $F_1$ on the remainder meets the target.  Subscriber counts are not available.}}
\end{tablenotes}
\end{threeparttable}
\end{table}%

\begin{table*}[!t]
\centering
\setlength\doublerulesep{0.5pt}
\caption{\review{Reproducibility settings.}}
\vspace*{-0.5em}
\label{tab:repro}
\begin{tabular}{p{0.2\textwidth}p{0.72\textwidth}}
\hhline{==}
\textbf{\review{Item}} & \textbf{\review{Setting}} \\
\hline
\review{Feature densification} & \review{Quantile transform of \eqref{eq:quantile-map} with $Q=5$ levels $(q_1,\dots,q_5)=(0,25,50,75,100)$; five-fold growth in the number of columns.} \\
\review{Cold subscribers} & \review{Maximum tenure below $30$ days: not scored, reported as inconclusive.} \\
\review{Missing values} & \review{Columns with more than $60\%$ missing values dropped; remaining missing values passed to XGBoost, which handles them natively.} \\
\review{Class balance} & \review{Balanced training dataset requested; \texttt{scale\_pos\_weight} set to the negative-to-positive ratio of the training split, which is about $1$ on the balanced data.} \\
\review{Data split} & \review{$30\%$ of subscribers for training (six months of history); the remaining $70\%$ unseen (Case~2); Cases~3-1 and~3-2 are later windows drawn from subscribers outside the training sample; inside the training sample, a $70/30$ split into training and cross-validation sets.} \\
\review{Model} & \review{XGBoost, \texttt{binary:logistic} objective, evaluation metric \texttt{aucpr}, random seed $42$.} \\
\review{Search procedure} & \review{Randomized search, $35$ sampled settings, stratified $3$-fold cross-validation; scoring is the $F_1$ score at the optimal threshold of each fold.} \\
\review{Search space} & \review{\texttt{n\_estimators} $\{200,350\}$; \texttt{max\_depth} $\{8,10,12\}$; \texttt{learning\_rate} $\{0.1,0.2\}$; \texttt{subsample} $\{0.6,0.8,1.0\}$; \texttt{colsample\_bytree} $\{0.4,0.6,0.8\}$; \texttt{min\_child\_weight} $\{1,3,5\}$; \texttt{gamma} $\{5,7,9\}$; \texttt{reg\_alpha} $\{0.3,1,10\}$; \texttt{reg\_lambda} $\{3,5,7\}$.} \\
\review{Truncation} & \review{Label-aware diagnostic: subscribers ranked by the probability assigned to their true class; largest subscriber $k^\star$ subset with $F_1\geq0.75$ retained, predictions at cutoff $0.5$.} \\
\review{Drift monitoring} & \review{PSI threshold $0.45$ per feature and null-ratio threshold $0.02$.} \\
\review{Explanations} &  \review{SHAP's log-odds scale.} \\
\review{Metrics} & \review{Precision, recall, $F_1$, precision--recall curves, and decile lift.} \\
\hhline{==}
\end{tabular}
\end{table*}%

\subsection{Value segments and value leakage}
Value segments, as mentioned earlier, are constructed using the primary churn predictors (e.g., tenure, expenditure, usage, and experience).

As discussed earlier, value segments are constructed using the primary predictors of churn, including tenure, expenditure, usage, and service experience. Each predictor is divided into a predefined number of levels, and the combination of these levels determines the total number of distinct value segments. For simplicity, we use three levels for subscriber expenditure (i.e., average revenue per user commonly known by the acronym ``ARPU''), three levels for tenure, two levels for usage, and two levels for service experience. The resulting combination produces $R = 36$ distinct value segments ($= 3 \times 3 \times 2 \times 2$) as per \eqref{eq:levels}.  \review{The levels of expenditure, tenure, usage, and experience were set by the operator's business rules and were not fitted to the churn labels.}  We depict the results in Fig.~\ref{fig:val_contr}, where these segments are ranked ordinally from highest to lowest value \review{(the ordering follows the level codes of the $3\times3\times2\times2$ structure)}. The top-ranked segments (e.g., Rank 1) exhibit high tenure and expenditure combined with optimal experience scores and relatively lower usage intensity; this profile correlates with high satisfaction, strong margin contribution, and high loyalty. Conversely, the bottom-ranked segments (e.g., Rank 36) display the opposite characteristics.

Fig.~\ref{fig:val_contr} illustrates the cumulative distribution of subscribers and churners. It demonstrates a clear inverse relationship between value segment ranks and their respective churn rates. As subscriber value declines, loyalty decreases and churn rates rise. We observe a distinct shift in the churn profile starting at the $20$th segment, where the cumulative churn rate begins to accelerate. This trend steepens sharply beyond the $30$th segment, indicating that the lowest-value cohorts are disproportionately driving the cumulative churn volume.%

In this context, we shed more light on value leakage as outlined in Section~\ref{sec:framework}.  Value leakage is the decline in the value contribution of retained subscribers who, while not churning, migrate to lower value segments. This erosion is typically driven by reduced expenditure, unmonetized usage increases, or degraded service experience. To quantify this, we analyze subscriber migration across two observation windows, explicitly excluding churners to isolate the behavior of the active base. Net value leakage is calculated as the differential between downgrade and upgrade rates within each segment.

Fig.~\ref{fig:val_leakage} reveals a concerning insight, high-value segments exhibit a higher propensity for value leakage. The cumulative net leakage rises steadily across the top four segments, stabilizes between segments 4 and 12, and then accelerates sharply between segments 12 and 16. However, an inflection occurs beyond segment 24, where cumulative leakage enters negative territory indicating net value formation, where retained subscribers are increasing their contributions. The strategic imperative is clear; we must deploy prescriptive interventions to address this value erosion specifically within the high-value tiers.  \review{These patterns are descriptive: they do not by themselves establish that any intervention would prevent churn, and causal retention recommendations require the uplift or treatment-effect analysis outlined in Section~\ref{sec:direction}.  No sensitivity analysis over alternative segment boundaries is run; thus, the patterns in Figs.~\ref{fig:val_contr} and~\ref{fig:val_leakage} are shown for this one segmentation.}
To address the value leakage observed in Fig.~\ref{fig:val_leakage}, prescriptive analytics and uplift analysis can be utilized to proactively influence subscriber trajectories.

Uplift analysis combined with recommendation system can be deployed to isolate the causal effect of specific campaign treatments \cite{fader2010customer,sato2019uplift,zhao2019uplift}, identifying subscribers who will respond positively to an intervention. By calculating the net incremental value of these targeted actions, the marketing platform can deploy precise treatments that maximize the probability of ``up movements'' (upgrades and retention) while significantly lowering the probability of ``down movements'' (churn and downgrades). Such a proactive intervention effectively stabilizes the high-value base, transforming the natural erosion of subscriber value into a profitable lifecycle management.

\begin{figure}[!t]
\centering
\includegraphics[width=0.48\textwidth]{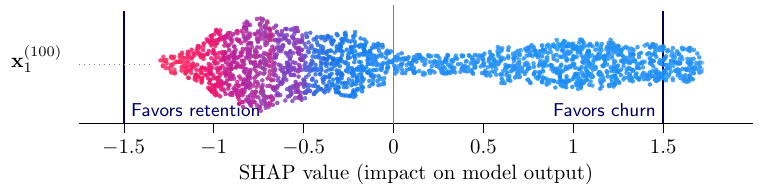}
\caption{Using SHAP beeswarm plot as an input for insights. \review{The horizontal axis is the SHAP value on the model's log-odds (margin) scale.}}
\label{fig:shapley_insight}
\end{figure}

\subsection{Insights}
In Fig.~\ref{fig:shapley_insight}, we observe a single result from the SHAP beeswarm plot.  This plot belongs to the transformed feature $\mathbf{x}_1^{(100)}$ which represents the maximum ($q_Q = 100$) subscriber tenure as measured in days (check Table~\ref{tab:meta}).  Blue (red) dots represent low (high) value of the maximum subscriber tenure.  Given there is a larger number of blue dots, we conclude that lower values of the maximum subscriber tenure dominate \review{the predicted churn flag}.  What remains is knowing the direction of the relationship the churn flag towards retention (i.e., $y = 0$) or attrition (i.e., $y = 1$).  In this case, it is clear that the direction of the low value of this feature is towards favoring churn.  A state diagram as shown in Fig.~\ref{fig:state_diag} explains the generation of an insight in plain English:

\fbox{\parbox{0.9\linewidth}{%
``Subscribers with higher~\textbar~lower \texttt{metadata}$(\mathbf{x}_i^{(Q)})$ tend to have higher~\textbar~lower churn risk.''
}}%

Here, \texttt{metadata}$(\mathbf{x}_i^{(Q)})$ refers to Table~\ref{tab:meta} and the choice between higher and lower is determined based on the rates of the dominance and relationship.


 
 Based on the top ten predicting features from the commercial dataset and the SHAP-based insights syntax outlined in our proposed framework, the framework generates one insight per top predictive feature, for a total of ten insights.  Here are three generated insights as a representative example:

\begin{itemize}
    \item Subscribers with lower maximum subscriber tenure (in days) tend to have higher churn risk.
    \item Subscribers with lower median revenue in the current month (in USDs) tend to have higher churn risk.
    \item \review{Subscribers with lower minimum quality of experience composite score tend to have higher churn risk.}
\end{itemize}%

\review{These insights describe predictive contributors, that is, features associated with higher predicted churn risk; whether optimizing for QoE or running retention and upsell campaigns would reduce churn requires the treatment-effect analysis mentioned in Section~\ref{sec:framework}.}
\newpage
\section{Limitations and Future Directions for Prescriptive Analytics}\label{sec:direction} 

\review{We start with listing limitations of this study: The dataset is proprietary and no longer available, which limits external reproducibility and comparison against other classifiers or against conventional aggregation.  Performance degrades across inference windows, and the precision drop in the second window is not explained by the drift indicators we monitored.  Truncation is a label-aware diagnostic, not a deployable selective-prediction rule, and $F_1$ on retained subscribers is not comparable to $F_1$ on all subscribers.  The model was trained on balanced classes.  SHAP attributions and value-leakage patterns are descriptive and not necessarily causal.  Results come from one operator and may not generalize to other operators or markets.}

Another area for future direction of this research line is to introduce prescriptive analytics.  This can be enabled through having the Business Operations Layer incorporate subscriber retention and value management. Since the Service and Resource Operations layers are in charge of services reconfiguration, the requirements to improve subscriber retention and value management need to be passed down on to them from the Business Operations Layer.  This is done in an agentic intent-driven autonomous fashion aligned with the AN reference architecture in Section~\ref{sec:autonomous}.  The interactive and agentic coordination mechanisms across AN layers is a challenging problem and requires specifications along many dimensions such as response times, inventory considerations, and---above all---cost and profitability objectives.  \review{Moreover, to quantify the impact of any given feature on churn, the ``feature elasticity'' as mentioned in Fig.~\ref{fig:mlarchitecture} can be derived and added to the insights to share a more quantitative measure of how sensitive the predicted churn propensity is to a marginal change in that feature (e.g., expressed as a percentage change in churn probability per percentage change in the feature's value).}


\vspace*{-1em}
\section{Conclusions}
This paper presented an explainable AI framework for predicting subscriber churn at telco scale.  This framework maps onto the TM Forum Autonomous Networks architecture through a feature densification transform that reconciles learning features of heterogeneous length. The framework met the recall and F\textsubscript{1} performance targets when applied to real dataset of tens of millions of subscribers, with SHAP-based explainability showing that, \review{in our data and feature set, commercial and experience features are stronger predictive contributors to churn than aggregated network counters despite the larger representation of the latter in the dataset.} \review{Our label-aware truncation diagnostic showed that $F_1$ on the retained subscribers meets the target after about $3$\% of subscribers, selected using the true labels, are set aside; it is not a deployable rule and is not comparable to $F_1$ on all subscribers.}  We further demonstrated how value segment analysis shows that churn risk rises as subscriber value declines, with high-value segments disproportionately exposed to leakage rather than outright attrition. Finally, we showed numerical results and a method to generate English-based insights based on the outcome of the proposed framework.


\bibliographystyle{IEEEtran}
\bibliography{main.bib}

\balance

\vspace*{-3em}
\begin{IEEEbiographynophoto}{Faris B. Mismar} received the B.S.\ degree with high distinction in Computer Engineering from the University of Jordan in 2004, the M.S.\ degree in Electrical Engineering and the MBA degree---both with highest distinction---from The University of Texas at Dallas in 2011 and 2014, respectively, and the Ph.D.\ degree in Electrical and Computer Engineering from The University of Texas at Austin in 2019.

From 2004 to 2008, he was a Telecommunications Engineer with Motorola. From 2009 to 2018, he held senior technical and leadership roles with Ericsson, including Senior Solutions Manager, Radio Access Networks. From 2019 to 2020, he was Director of Radio Performance Assurance and Engineering with Samsung Electronics America. He joined Nokia in 2020 as Head of Data Science and Analytics Practice, Mobile Networks. Since 2022, he has been a Distinguished Member of Technical Staff and Senior Principal Consultant with Nokia Bell Labs Consulting, Dallas, TX, USA. He joined the Electrical and Computer Engineering Department, University of Texas at Dallas, as an Assistant Adjunct Professor in 2021 and has served as an Associate Adjunct Professor since 2024.

An accomplished technical leader and innovator, his research has directly influenced 3GPP conditional handovers procedures (3GPP TS 38.300) and facilitated the creation of production O-RAN rApps and xApps.  In his career, he has led or contributed to over $100$ large-scale client-facing projects worldwide including a lunar surface communications infrastructure study.  Dr.\ Mismar received the University of Jordan Award for Excellence in 2004 and the Marcus Wallenberg Foundation Scholarship for Scientific Research and Education in 2016 and 2017. He received the Nokia Bell Labs Distinguished Member of Technical Staff Award in 2023 (top $0.3\%$) for ``sustained contributions with proven impact'' and excellence in telecommunications and AI. He is a Senior Member of the IEEE, served as an Associate Editor of \textsc{IEEE Transactions on Machine Learning in Communications and Networking} since 2025, and is a ComSoc Distinguished Lecturer (2026-2027).  His research interests include machine learning, network optimization, and wireless communications.
  

\end{IEEEbiographynophoto}

\begin{IEEEbiographynophoto}{Abdol Saleh} is a Principal Consultant and a Distinguished Member of Technical Staff at Nokia Bell Labs. With over three decades of experience in the telecommunications industry, he specializes in digitalization technologies, AI/ML and big data platforms. His expertise includes network automation, industrial and telecommunication econometrics analysis, customer experience and value management. Abdol holds a PhD in Industrial Engineering and Operations Research from Lehigh University (1988). He has continued to expand his expertise with advanced certifications, including a Professional Certificate in Data Engineering from MIT (2024), an Applied Data Science Certificate from MIT (2022), and a Smart Industry Readiness Index CSA Certification (INCIT, 2021), reinforcing his commitment to driving digital transformation in telecommunications and beyond.
\end{IEEEbiographynophoto}

\begin{IEEEbiographynophoto}{Ivan Maxmillian Putra Pasaribu}
is a Senior Officer of Data Science in the Data Analytics and AI division at a telecommunications operator in Indonesia. His work focuses on the application of machine learning and artificial intelligence to network analytics, customer experience management, and business process optimization in telecommunications. He has led AI use case implementation and cross-functional collaboration to advance AI adoption across the company. Ivan holds a Bachelor of Engineering in Electrical and Electronics Engineering from the University of Indonesia (2023). His research interests include machine learning, data analytics, and network optimization.
\end{IEEEbiographynophoto}

\begin{IEEEbiographynophoto}{Suhelmy Syaifuddin}
is the Vice President and Head of Data Analytics and AI/ML at a leading telecommunications operator in Indonesia. With over two decades of experience spearheading transformative initiatives in the telecommunications and data sectors, his leadership focuses on the end-to-end development of scalable agentic AI solutions and customer experience excellence. His professional background includes the development of diverse SaaS products, alongside the management of large-scale IT and analytic projects. His current research and professional interests are centered on scalable machine learning and agentic customer experience intelligence.
\end{IEEEbiographynophoto}

\end{document}